\documentclass[conference]{IEEEtran}
\IEEEoverridecommandlockouts
\usepackage{balance}
\usepackage{siunitx}
\usepackage{custom}
\usepackage{multirow}

\begin{document}
\title{Eco-Routing Using Open Street Maps}
\author[1]{R. K. Ghosh}
\author[2]{Vinay R}
\author[3]{Arnab Bhattacharya}

\affil[1]{Department of EECS, IIT Bhilai, Raipur 492015, INDIA\\ Email:{\tt rkg@iitbhilai.ac.in}}
\affil[2,3]{Department of CSE, IIT Kanpur, INDIA\\ Email: {\tt \{vinayr, arnabb\}@cse.iitk.ac.in} \vspace{1.5ex}}

\maketitle

    \begin{abstract}
  
A vehicle's fuel consumption depends on its type, the speed, the condition, and the gradients of the road on which it is moving.  We developed a Routing Engine for finding an eco-route (one with low fuel consumption) between a source and a destination. Open Street Maps has data on road conditions. We used CGIAR-CSI road elevation data~\cite{srtm_ascii_tiff} to integrate the road gradients into the proposed route-finding algorithm that modifies Open Street Routing Machine (OSRM). It allowed us to dynamically predict a vehicle's velocity, considering both the conditions and the road segment's slope. Using Highway EneRgy Assessment (HERA) methodology, we calculated the fuel consumed by a vehicle given its type and velocity. We have created both web and mobile interfaces through which users can specify Geo coordinates or human-readable addresses of a source and a destination. The user interface graphically displays the route obtained from the proposed Routing Engine with a detailed travel itinerary.
\end{abstract}
	\section{Introduction}
\label{ch:int}

The stiffest challenge faced by humankind today is to maintain a low ambient Air Quality Index (AQI)~\cite{MASIOL201484}. Besides having an adverse impact on climate~\cite{challengeAQI,medical2016,nationalGeo2019}, poor air quality affects flora and fauna. The presence of chemicals, biological, and physical particles in the atmosphere is the major reason for air pollution. 
It is not possible to prevent naturogenic pollution. However, it remains under the tolerance limit over time. The pollution due to anthropogenic sources increases in proportion due to accelerated human-activities~\cite{Grimm756}. The significant causes of pollution are toxic emissions from industries, agricultural activities, waste accumulation, and fossil fuel combustion in the generation of electricity and transportation service. 

\begin{figure}[htb]
  \begin{center}
    \centering
    \includegraphics[scale=0.4]{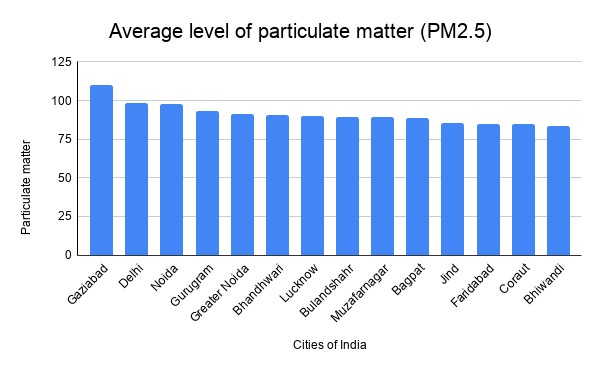}
    \caption{Contribution of the transport sector to total emissions of the main air pollutants~\cite{transport}.}
    \label{fig:pmLevel}
  \end{center}
\end{figure}

The emission due to the burning of fuel in the transportation sector is a dominant factor affecting air quality, especially in large cities. It releases Green House Gases (GHG) and Non-Methane Volatile Organic Compounds (NMVOC) into the atmosphere. NMVOC mainly contributes to particulate matter. Controlling particulate matter is a significant challenge especially in India. Transportation sources lead to almost one-third of PM pollution in India. In fact, ten most polluted cities in the world are located in India. Figure~\ref{fig:pmLevel}~\cite{statista2019} gives an idea of the severity levels of PM pollution in different cities across northern India.

\begin{figure}[htb]
  \begin{center}
    \centering
    \includegraphics[scale=0.25]{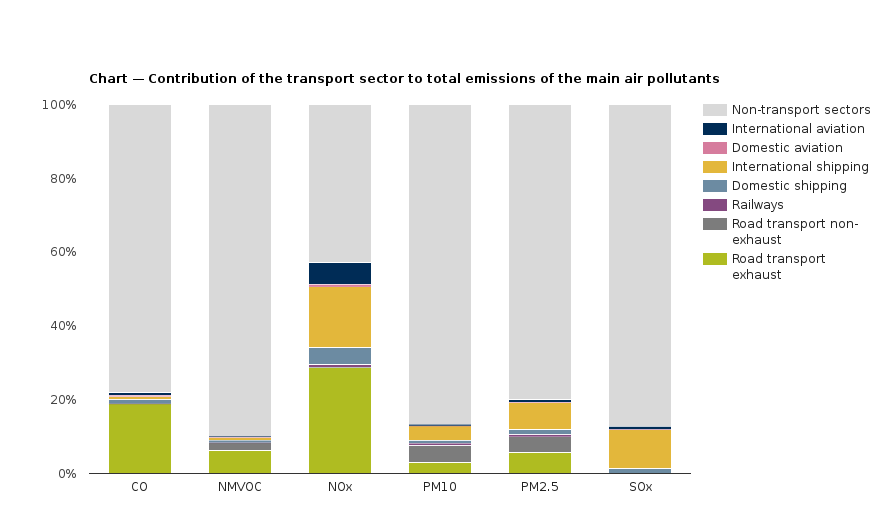}
    \caption{Contribution of the transport sector to total emissions of the main air pollutants~\cite{transport}.}
    \label{fig:pollutants}
  \end{center}
\end{figure}

\begin{figure}[htb]
  \begin{center}
    \centering
    \includegraphics[scale=0.4]{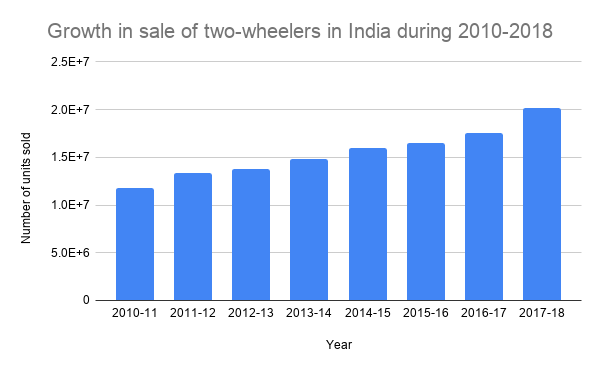}
    \caption{Two wheeler sales in India during the period 2011-2018.}
    \label{fig:twowheelersales}
  \end{center}
\end{figure}

The composition of direct GHG is water vapor (H$_2$O), Nitrous Oxide (NO), Carbon Dioxide (C$_2$O), Methane (CH$_4$), Ozone (O$_3$). Indirect GHG has NMVOC, Sulpher Dioxide (S$_2$O), and Carbon Monoxide (CO) in its composition. Without GHG, the average surface temperature of Earth would be -18$\si{\degree}$ instead of the present average of 15$\si{\degree}$~\cite{le2005historical}. However, excessive emission of indirect GHG may cause a significant rise in global temperature and adversely affecting the Earth's climate~\cite{RAMANATHAN200937}. 
As one can observe from the Fig~\ref{fig:pollutants}, the most significant factor contributing to emissions of all the major air pollutants except SO$_2$ happens to be road transportation system. 
The severity of pollution in Indian cities has created a near-disastrous situation for human survival. The economic growth coupled with population growth in cities has also led to increased mobility. The number of registered vehicles in India has multiplied in the last decade and is still growing. According to a report~\cite{wagner2019}, during the eight years 2011-2018, the increase in sales of two-wheelers alone is from 117 million to 210 million as indicated by Fig.~\ref{fig:twowheelersales}. 
Not only does the fuel used in the residential and transportation sectors contribute to the air pollutants,  but the movements of vehicles also cause re-suspension of dust, keeping the density of the ambient concentrations of particulate matter more or less at the same level. 

The minimum travel time is the most important optimization in measuring the efficiency of transportation services.  Geographic Information systems (GIS) help in achieving it. All the web/mobile GIS systems provide the shortest distance or the shortest time routes by collecting real-time or cumulative traffic patterns. The fuel consumption of a vehicle is not a linear function of velocity. So, optimizing the route for the shortest possible time does not necessarily optimize a vehicle's fuel consumption. It is essential to work on the tools that reduce emissions from transportation services. The cumulative effects of creating such a tool-chain can be summarized as follows:
\begin{itemize}
    \item It lowers the cost of transport and contributes to
    the overall fiscal management of a nation's economy.
    \item It reduces the emission improves AQI in cities, making them more livable. Overall impacts lead to a better environment.
    \item It leads to the conservation of fossil fuel, which is estimated to last only between 50-100 years at the present rate of consumption. 
\end{itemize}

This paper provides a method for estimating fuel consumed by the vehicle for a given path. By incorporating the estimation of fuel with the technique of finding the shortest route between a pair of source and destination, we can estimate the fuel paths for traveling on vehicles on the roads. We have developed a web and a mobile interface to find a complete travel itinerary by considering the fuel consumption factor.

    \section{Background and Related Work}
\label{ch:bcg}

Eco-friendly routing of vehicular movements is one of the significant aspects of the overall approach for control of emission and conservation of fossil fuel. Before addressing the problem, we need to understand the background of appropriate models used to estimate emission and fuel consumption for different vehicle types. The concern here is to integrate both fuel consumption assessments and the road elevation data with widely use routing machines such as Google Maps or Open Street Maps to generate an eco-friendly route between a source and a destination. 

\subsection{Eco-friendly Routing}

Petroleum-based fuel accounts for almost 35\% of the input cost in the transportation industry~\cite{gohari2018effects,litman2009transportation}. Any fuel-saving measures will be a boon for the industry and cut down the emissions drastically. Eco-routing solutions~\cite{SCHRODER201940} allow for a prediction of the most fuel-economic routes. 

Many factors affect vehicle emissions and fuel consumption. However, the three most important factors are vehicle characteristics, road characteristics, and traffic conditions. Each type of vehicle has different aerodynamic drags, different engine technology, which determines its fuel efficiency. Fuel consumption and vehicle emissions are sensitive to changes in travel speed, traffic patterns, etc. A small portion of the route may account for a substantial part of total fuel consumption depending on traffic intensity~\cite{wang2019real}. The gradient of a road plays another important role in understanding a vehicle's fuel efficiency as real-world fuel consumption and emission models are sensitive to the road grade estimates. Some research~\cite{yao2013study} indicates that the road grade accounts for approximately 1-3\% of fuel consumption used in light vehicles. The eco-friendly route will be the least fuel-consuming route among all the possible routes given a source and a destination. Therefore, the estimate of fuel consumption requires, so we need the entire road curvature data for the complete path from a source to a destination. Finally, it is extremely convenient and a practical aid for a vehicle driver if Petroleum-based fuel accounts for almost 35\% of the input cost in the transportation industry~\cite{gohari2018effects,litman2009transportation}. Any fuel-saving measures will be a boon for the industry and cut down the emissions drastically. Eco-routing solutions~\cite{SCHRODER201940} allow for a prediction of the most fuel-economic routes. 

Many factors affect vehicle emissions and fuel consumption. However, the three most important factors are vehicle characteristics, road characteristics, and traffic conditions. Each type of vehicle has different aerodynamic drags, different engine technology, which determines its fuel efficiency. Fuel consumption and vehicle emissions are sensitive to changes in travel speed, traffic patterns, etc. A small portion of the route may account for a substantial part of total fuel consumption depending on traffic intensity~\cite{wang2019real}. The gradient of a road plays another important role in understanding a vehicle's fuel efficiency as real-world fuel consumption and emission models are sensitive to the road grade estimates. Some research~\cite{yao2013study} indicates that the road grade accounts for approximately 1-3\% of fuel consumption used in light vehicles. The eco-friendly route will be the least fuel-consuming route among all the possible routes, given a source and destination. Therefore, the estimate of fuel consumption requires, so we need the entire road curvature data for the complete path from a source to a destination. Finally, it is extremely convenient and a practical approach for a vehicle driver if the existing navigation tools may provide eco-route for commutation and transportation. Finding eco-friendly routing between a source and a destination requires three basic components:
\begin{itemize}
    \item A model for estimating emissions from different vehicle types and a compilation method for calculating aggregated emission data under different road surfaces and traffic conditions.
    \item A database of roads providing a digital road map with historical data and data fusion strategies for integrating crowdsourced real-time traffic.
    \item A user interface for taking input and display the route on a road map, along with fuel consumed, and travel itinerary 
\end{itemize}

Developing an eco-friendly navigation tool is far too complex to build from scratch. It requires a sustained long term large team effort, which can only be possible by Google, Microsoft, etc. Therefore, in this paper, we employed the following short cuts:
\begin{enumerate}
    \item For road network databases, we leveraged two well-established navigation and route machines, namely, Google Maps~\cite{GoogleMaps} and Open Street Maps~\cite{OSM}.
    \item For estimating fuel consumption, we used HERA methodology~\cite{HERA2016}.
    \item  For display of maps and least fuel cost routes, we used a third-party front-end builder. 
    \item For compiling travel itinerary, we used Google multipoint representation of turns, etc. 
    \item For integrating a real-time mobile navigation interface, we employed another third party application. 
\end{enumerate}

\subsection{HERA Methodology}

Many models exist for the estimation of emissions due to road traffic and energy consumption.  We can analyze the traffic flows from two general perspectives,
\begin{enumerate}
    \item Microscopic.
    \item Macroscopic.
\end{enumerate} 
A Microscopic view of traffic takes into account the vehicle-related variables and traffic flow characteristics. The vehicle-related variables, among others, include the length of the vehicle, the speed, and the acceleration. The traffic flow characteristics refer to the variability in traffic streams, which depends on physical constraints and the complex behavior of driving. The distance between two vehicles and time variability between different traffic streams constitutes variability in a traffic stream. For macroscopic analysis of traffic, we try to extrapolate an aggregate picture of vehicular movements~\cite{trafficflow} instead of looking into each vehicle in a traffic stream.

Two prominent simulators widely used to model emission from vehicles are: 
\begin{itemize}
    \item The COPERT~\cite{COPERT2009} – COmputer Programme to calculate Emissions from Road Transport.
    \item  MOVES~\cite{MOVES2010} – Motor Vehicles Emission Simulator.
\end{itemize}
Both models use their databases to obtain the emission factors as a function of the average cycle of speeds. COPERT is mainly used in Europe, while MOVES is used in the USA. Other types of simulators take microscopic details into account to estimate emission maps with speed and acceleration.

The energy consumption and emission modeling tools provide quantitative evaluation for a given route.
For our purposes, we have used models created based on HERA~\cite{HERA2016} methodology. It uses an average speed consumption model based on COPERT emission factors.  These factors are adjusted with a correction factor to incorporate the vehicles' gradient effect in each road section.

The average speed is a key factor in fuel consumption and emissions of a vehicle. Several studies~\cite{HERA2016} have found a relationship between fuel consumption and vehicle speed. Therefore, it is very appropriate to use models which are based on speed. Since the HERA consumption model is based on consumption and emission factors from COPERT, it considers a set of vehicle-centric parameters to calculate fuel consumption. For example, it uses the fuel type and engine capacity, and engine technology for light vehicles (passenger cars).  It uses the maximum total weight, load state, engine technology, and road gradient for heavy vehicles.

HERA~\cite{HERA2016} uses a slightly altered set of parameters. It take the average speed as the leading factor and the road gradient as a correction factor. It considers six vehicle types, namely,  passenger cars, light-duty vehicles, motorcycles, rigid trucks, articulated trucks, and buses.  HERA methodology provides an analytical formula to determine an approximates relationship between the fuel Consumption and the average speed for each category of vehicle. Table~\ref{tab:tab1background} summarizes the relationships.
\begin{table*}[htb]
\begin{center}
  \begin{tabular}{| m{5cm} | m{5cm} |}
    \hline
      \multicolumn{1}{|c|}{\textbf{Vehicle Type}} & \multicolumn{1}{c|}{\textbf{Fuel Consumption Function for 0\% Road Gradient (gfuel/veh-km)}} \\
    \hline\hline
      \multicolumn{1}{|c|}{Motor Cycle} & \multicolumn{1}{c|}{25.722 + (276.13/V) + (-0.254)*V + 0.00311*V$^2$} \\
    \hline
      \multicolumn{1}{|c|}{Passenger Car} & \multicolumn{1}{c|}{54.7 + (496/V) + (-0.542)*V + 0.0042*V$^2$} \\
    \hline
      \multicolumn{1}{|c|}{Light-Duty Vehicle} & \multicolumn{1}{c|}{146.27 + ((-0.0000106)/V) + (-2.596)*V + 0.01984*V$^2$} \\
    \hline
      \multicolumn{1}{|c|}{Rigid Truck} & \multicolumn{1}{c|}{152.96 + (604.156/V) + (-2.295)*V + 0.0238*V$^2$} \\
    \hline
      \multicolumn{1}{|c|}{Articulated Truck} & \multicolumn{1}{c|}{332.603 + (1680.879/V) + (-4.676)*V + 0.0311*V$^2$} \\
    \hline
      \multicolumn{1}{|c|}{Bus} & \multicolumn{1}{c|}{281.735 + (4186.178/V) + (-3.457)*V + 0.0216*V$^2$} \\
    \hline
  \end{tabular}
\end{center}
\caption{Approximate relationships between speed and fuel consumption for each vehicle type.}
    \label{tab:tab1background}
\end{table*}
Overall Fuel Consumption, 
\[
     FC_i =  FCZ_i * RGF_i,
    \label{eq:FC}
\]
where $FCZ$ is the fuel consumption on the road with zero gradients, $RGF$ is the Road Gradient Factor, and $i$ represents a different vehicle type such as a car, bus, etc.

HERA uses each vehicle category's consumption factors and consumption curves data from COPERT database to construct the above formulae.  By applying the non-linear regression techniques and parametrization of the consumption curves, HERA derives the consumption model function for each vehicle type, excluding the Road Gradient Factor RGF.

\subsection{Road Gradient Factor}

Apart from average vehicle speeds, many other parameters affect fuel consumption. These include a wear-and-tear of vehicle, the technology used, and the conditions of the road, the elevation of the road (or Road Gradient), Road Surface.

The gradient of a road increases or decreases the resistance of a vehicle to traction. The increase or decrease in the engine load has a corresponding effect on emission and fuel consumption rates. The effect of road gradient on the vehicle is not symmetrical, i.e., the extra fuel consumed by traveling uphill won’t be equal to the reduced fuel consumption when going downhill. In general, the road gradient effect is more on the lighter vehicle than on the heavy-duty vehicles.

For a given speed $V$, the fuel consumed on the flat road is different compared to either uphill or downhill. HERA uses a correction factor which is based on the results of the German Emission Factor Program~\cite{de2004modelling}. Some of the factors included in the correction are:
\begin{enumerate}
  \item The road gradient
  \item The pollutant
  \item The average speed of the vehicle
  \item The vehicle category
\end{enumerate}
The road gradient factor $RGF_i$ for each vehicle type $i$ is found according to following formula:
\[
\begin{split}
RGF_i &= A6_{i,k} * V^6 + A5_{i,k} * V^5 + A4_{i,k} * V^4 + A3_{i,k} * V^3 \\
& + A2_{i,k} * V^2 + A1_{i,k} * V + A0, 
\end{split}
\]
where RGF$_i$ is the road gradient factor for the vehicle category $i$, $V$ is the vehicle's average speed in km/hr. $A0$ to $A6$ are the coefficients depending on the gradient and the type of vehicle provided by Hickman et al.~\cite{MEET1999}.
    \section{Google Maps and Open Street Map}
\label{ch:gmap}
Developing an eco-friendly vehicle routing from scratch is not feasible without the support of a large team. We aim to assess problems likely to appear in trying to enhance an existing navigation tool for eco-friendliness. Therefore, we started by exploring the different APIs provided by Google Maps and Open Street Map.

\subsection{Geo-Coding}
Let us first start with a precise definition of the problem. 
\begin{definition}[Problem statement] 
Given a source $S$ and a destination $D$, find the most fuel-efficient route from $S$ to $D$.
\end{definition}
For navigation, a user may specify a source or a destination. The source or the destination would be in the form of a human-readable address such as a street address. 
A human-readable street address is converted into geographic coordinates by a process known as Geo-coding. More precisely, given a street address as input, Geo-coding outputs a (latitude, longitude) pair. The geographic coordinates are convenient as they can be used to place markers on a map and find the shortest distance or elevation between a source and a destination. 

\subsection{Google Map APIs}
Google Maps API tool provides an API for extracting geo-coordinates for an address by using the following url:

\url{https://maps.googleapis.com/maps/api/geocode/json?address=<ADDRESS>&key=<KEY>}

Along with latitude and longitude coordinates, it also provides other address components such as state, country, pin code, etc.
For example, using the ADDRESS = "cse IIT Kanpur" returns Geo-coordinates (26.5143, 80.2348).

\subsubsection{Places Nearby}

Google API can be used to find important utility points near a pre-specified location. For example,  utility points of interest could be the nearby restaurants, car service stations, hospitals, fuel pumps, convenience stores, govt offices, etc. The API call given below extracts the location and other information about the nearby restaurants.  

\url{https://maps.googleapis.com/maps/api/place/nearbysearch/json?location=26.5143, 80.2348&rankby=distance&type=restaurant&key=<KEY>}\\
The above example returns the restaurant details near the given coordinates (CSE IIT Kanpur), sorted by the distance.

\subsubsection{Elevation}

The elevation values represent the elevations relative to the sea level. For finding the elevation at some Geo coordinates, we use the following GoogleMap API.

\url{https://maps.googleapis.com/maps/api/elevation/json?locations=26.5143, 80.2348&key=<KEY>}\\
It returns the elevation at CSE IIT Kanpur.

\subsubsection{Direction}

A series of Geo coordinates along the path represents a direction. To extract the Geo coordinate series for a given source-destination pair, we use the following GoogleMap API. We need to go via these Geo coordinates while driving.

\url{https://maps.googleapis.com/maps/api/directions/json?origin=<Geo Cordinates>&destination=<Geo Cordinates>&key=<KEY>}\\
This API returns polylines (a sequence of straight lines) from a source to a destination. This polyline contains encrypted geo coordinates. We need to decrypt them and extract a series of geo-coordinates from all those polylines.

\begin{figure*}[htb]
  \begin{center}
    \includegraphics[scale=0.4]{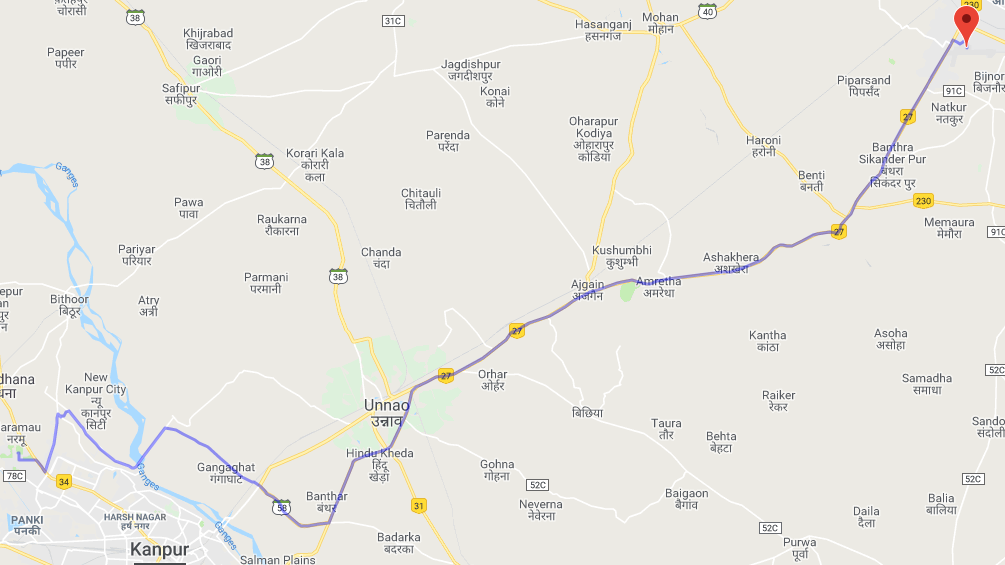}
  \end{center}
    \caption{Direction service from Google for route between CSE IIT Kanpur to Lucknow Airport}
    \label{fig:direction_gmap}
\end{figure*}

Figure~\ref{fig:direction_gmap} shows the graphical display of the direction from CSE IIT Kanpur to Lucknow Airport.

\subsubsection{Elevation Along a Path}
\begin{figure*}[htb]
  \begin{center}
    \includegraphics[scale=0.25]{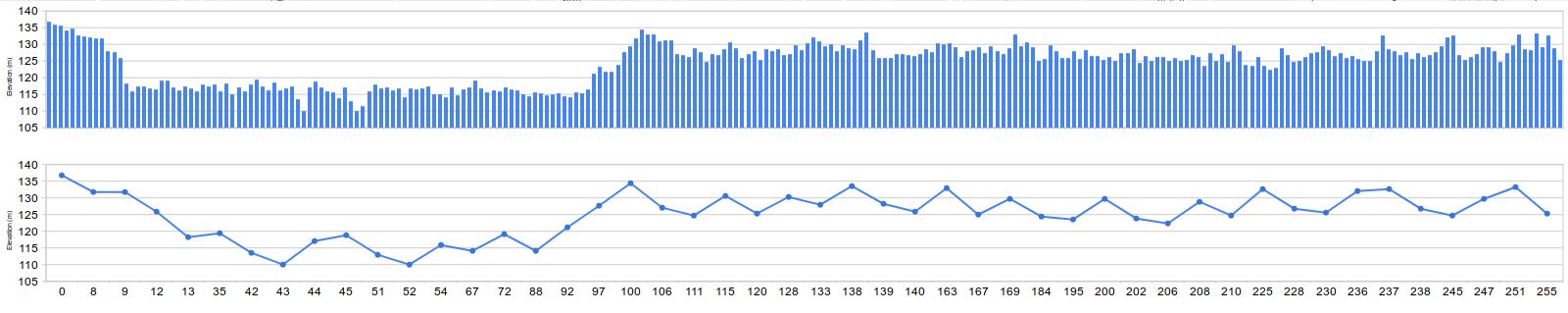}
  \end{center}
    \caption{Elevation and normalized elevation along the path}
    \label{fig:elevation_path}
\end{figure*}

By using `getElevationAlongPath' API, We can find the road elevations along the path from a source to a destination. For example the top half of Fig.~\ref{fig:elevation_path} shows the elevations along the path from ‘CSE IIT Kanpur’ to ‘Lucknow Airport’.

To understand how the road is, we normalized the road elevations by talking about the minimum raise/drop in the elevation between two consecutive points $ \triangle $ e=5m. The lower half of Fig.~\ref{fig:elevation_path} shows the normalized elevations.

Using the direction service, we can even find the distance and the time required to travel along the path. Therefore, it is possible to calculate the average velocity along the route. By using elevation service, we can calculate the gradient along a road. Once the road gradient is available, the equation~\ref{eq:FC} can be used to find the fuel consumption for a given route.
\subsubsection{Fuel Estimation}
\begin{figure*}[htb]
  \begin{center}
    \includegraphics[scale=0.2]{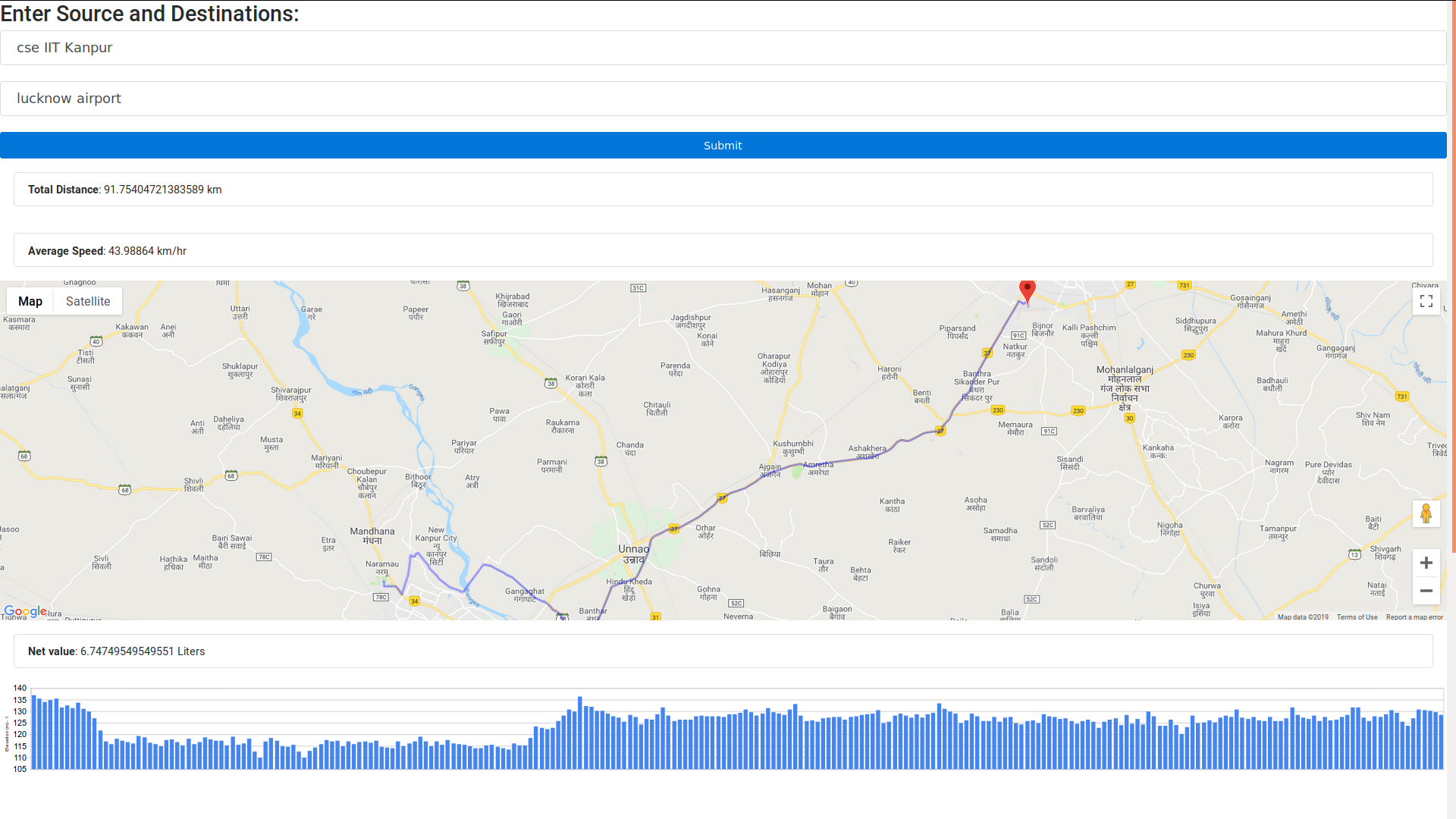}
  \end{center}
    \caption{CSE IIT Kanpur to Lucknow airport}
    \label{fig:fuel_gmap_1}
\end{figure*}

\begin{figure*}[htb]
  \begin{center}
    \centering
    \includegraphics[scale=0.25]{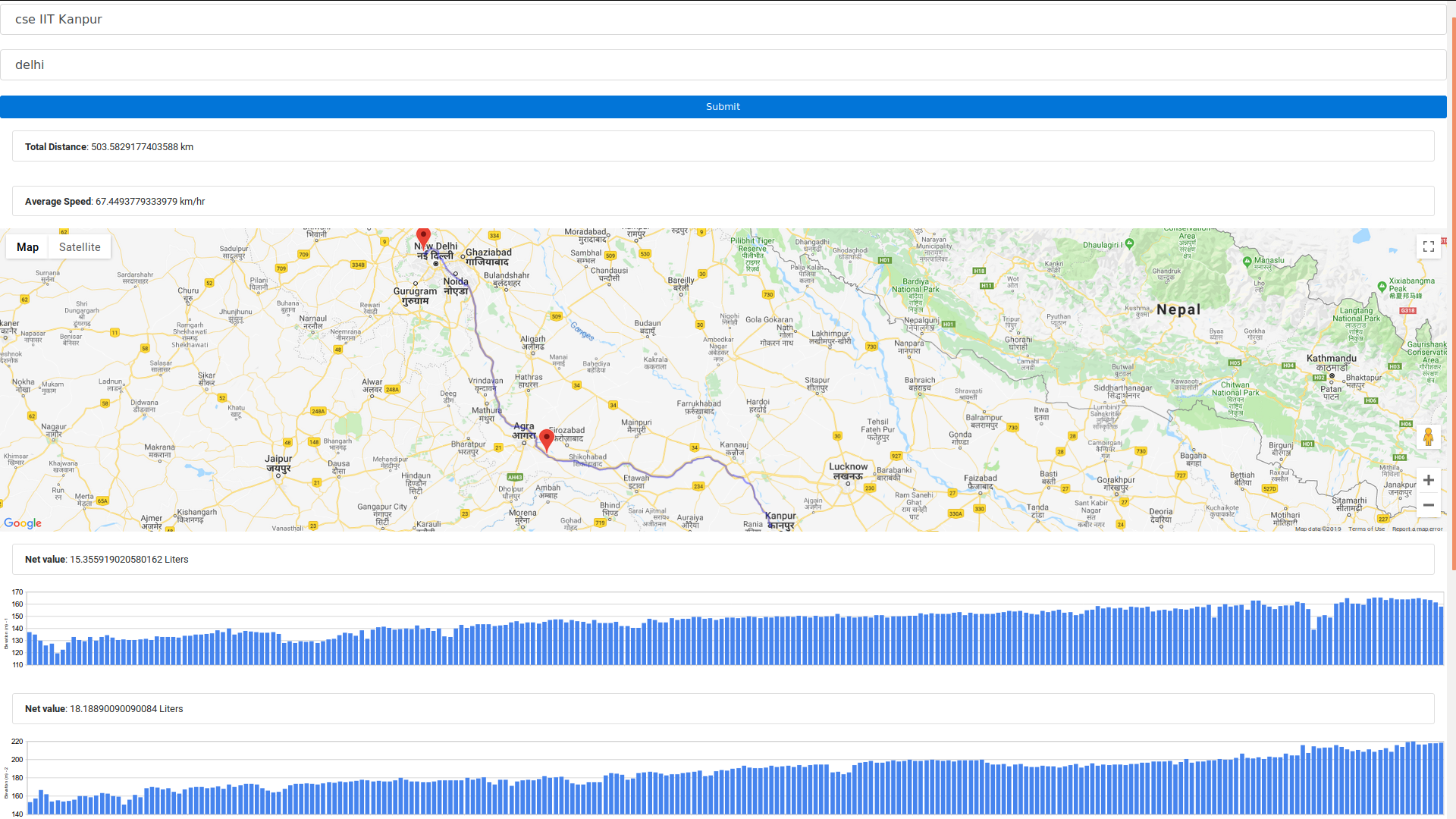}
    \caption{IITK to Delhi}
    \label{fig:fuel_gmap_2}
  \end{center}
\end{figure*}

We built a simple application that uses the above APIs and calculates Fuel consumption for the route provided by Google Maps. Fig.~\ref{fig:fuel_gmap_1} and~\ref{fig:fuel_gmap_2} provide screenshots of two different source-destination pairs obtained from our application.

We divided the road into intervals of 256km each to get a clear understanding of the road's elevation. 
As shown in the Fig.~\ref{fig:fuel_gmap_2} the first 256km has is a 20m rise in elevation incurring a fuel consumption of 14.5 liters. However, for the next part 247km, there is a 65m rise in the elevation, and the fuel consumption increases to 18.2 liters. So, the gradient of a road segment affects fuel consumption. A vehicle requires more fuel to travel on a steep ascent than a relatively lower ascent.

\subsection{Limitations of Google Map API}
The limitation of Google Maps is that it only estimates the fuel consumption for the given routes. In other words, it is possible to calculate fuel consumption if we already know the path. However, we aim to estimate the least fuel consumption route. So the routing algorithm has to incorporate fuel consumption as a basis for finding the path and not the time or distance. Unfortunately, Google Maps don't have an API which uses fuel consumption as a basis for route computation. Furthermore, we cannot apply our routing algorithms on Google Maps because it does not provide us the maps' raw data.

\section{Open Street Maps (OSM)}
Limitations of Google Map APIs forced us to search for the alternative to find raw map data. We searched the projects where the users collect the raw data. We found the open-source collaborative mapping~\cite{goodchild2007citizens} called "Open Street Map" (OSM) which not only competes with commercial services like
Google Maps but also provides the raw map data. OSM data is entirely crowdsourced, and anyone can use/edit the map data~\cite{OSM} at no cost at all. Structurally, OSM consists of two different types of elements:
\begin{itemize}
    \item Node: It contains places defined with it's Geo Coordinates
    \item Way: It contains the roads, boundaries of the building, places. The roads have a detailed description like the road's name, whether the road is motorway/trunk/residential, etc.
\end{itemize}

There is a project called Open Source Routing Machine (OSRM). It is a Routing Engine relying on the implementation of multi-level Dijkstra's shortest path algorithm on a graph built from the OSM data. The edge weights of the graph are distances or duration for covering those distances. Therefore, the routing based on the least distance or the least travel time. However, the advantage of the above implementation is that it allows one to fix a velocity, depending on the road type. Table~\ref{tab:OSRM_velocity} gives a specification of those values.
\begin{table}{htb}
\begin{center}
  \begin{tabular}{ | m{5cm} | m{5cm} | }
    \hline
      \multicolumn{1}{|c|}{\textbf{Road Type}} & \multicolumn{1}{c|}{\textbf{velocity(km/hr)}} \\
    \hline\hline
      \multicolumn{1}{|c|}{motorway} & \multicolumn{1}{c|}{90} \\
    \hline
      \multicolumn{1}{|c|}{motorway\_link} & \multicolumn{1}{c|}{45} \\
    \hline
      \multicolumn{1}{|c|}{trunk} & \multicolumn{1}{c|}{85} \\
    \hline
      \multicolumn{1}{|c|}{trunk\_link} & \multicolumn{1}{c|}{40} \\
    \hline
      \multicolumn{1}{|c|}{primary} & \multicolumn{1}{c|}{65} \\
    \hline
      \multicolumn{1}{|c|}{primary\_link} & \multicolumn{1}{c|}{30} \\
    \hline
      \multicolumn{1}{|c|}{secondary} & \multicolumn{1}{c|}{55} \\
    \hline
      \multicolumn{1}{|c|}{secondary\_link} & \multicolumn{1}{c|}{25} \\
    \hline
      \multicolumn{1}{|c|}{tertiary} & \multicolumn{1}{c|}{40} \\
    \hline
      \multicolumn{1}{|c|}{tertiary\_link} & \multicolumn{1}{c|}{20} \\
    \hline
      \multicolumn{1}{|c|}{unclassified} & \multicolumn{1}{c|}{25} \\
    \hline
      \multicolumn{1}{|c|}{residential} & \multicolumn{1}{c|}{25} \\
    \hline
      \multicolumn{1}{|c|}{living\_street} & \multicolumn{1}{c|}{10} \\
    \hline
      \multicolumn{1}{|c|}{service} & \multicolumn{1}{c|}{15} \\
    \hline
  \end{tabular}
\end{center}
\caption{Velocity on various roads}
    \label{tab:OSRM_velocity}
\end{table}
Because OSRM provides raw map data and the fact that it has a specification of vehicle velocity based on road type, we decided to implement our project on Open Street Maps and OSRM. 
    \section{Optimal Fuel Routing Machine (OFRM)}
\label{ch:ofrm}

The road network graph built from OSM by OSRM has edge weights assigned either as distance or as the time to those distances considering the road types.  For computing the least fuel cost path, we need to change the edge weights to fuel consumption parameters using equation~\ref{eq:FC}, which is straightforward. However, in the equation, we still have to incorporate  Road Gradient Factor. But the OSM data set like Google Maps doesn't have those details. So, the elevation needs to be computed only from available data. Let us find a method to determine the road gradient by integrating the elevation data.

\subsection{Integrating Elevation}

The gradient of a road affects the velocity of a vehicle moving on the way. Since speed affects a vehicle's fuel consumption, we need to know the road gradient to estimate fuel consumption when a vehicle moves either up or down a road gradient. We assign a penalty coefficient to velocity, which varies according to the known gradient of the road. The penalty function is determined based on the time required to cover a distance over flat road versus the time needed to cover the same distance on the road with known elevation either in ascent or in descent. A road elevation has the effect of slowing down the vehicle while moving up or down the gradient. Hence, the estimation of fuel consumption can be made based on the modified velocity of the car.  

\begin{figure}
    \centering
    \includegraphics[scale=0.2,angle=-90]{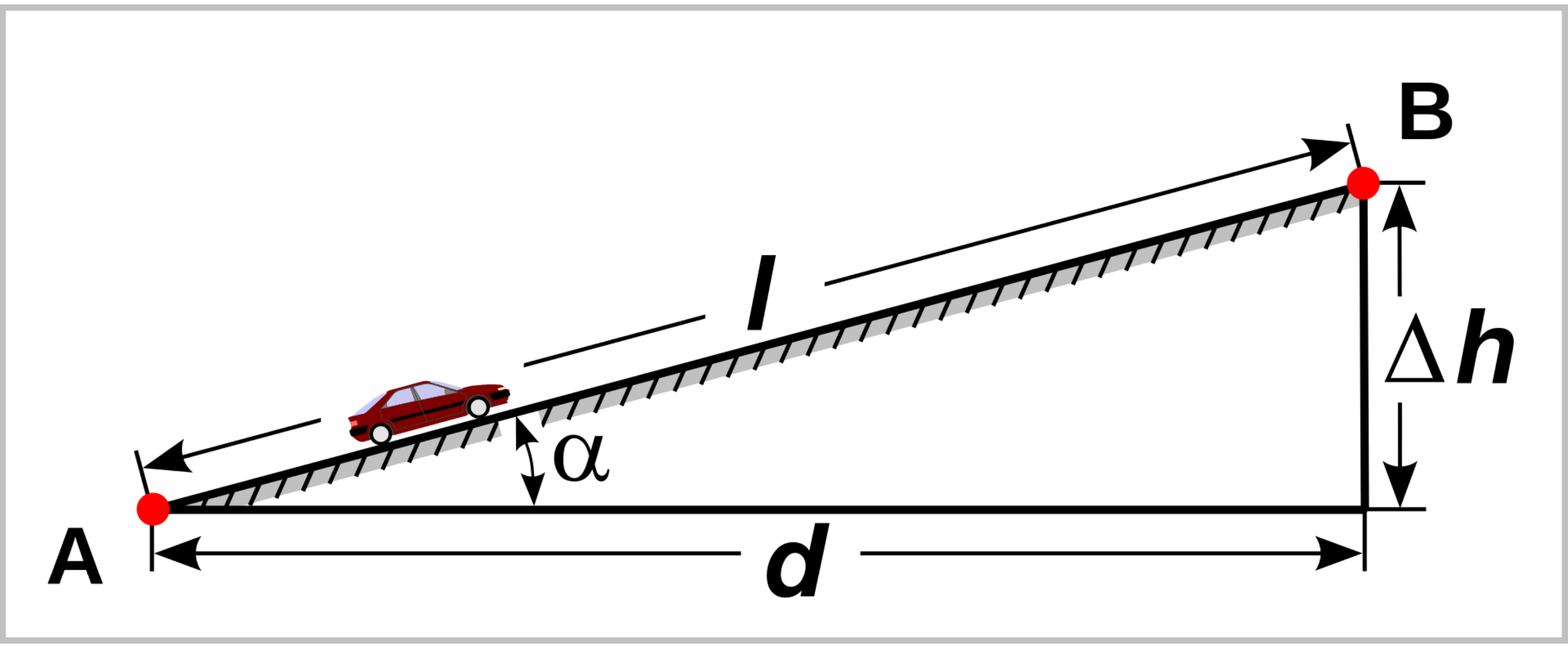}
    \caption{Gradient of a road segment.}
    \label{fig:altitude}
\end{figure}

Before deciding on a penalty factor, it is important to understand the road gradient's calculation as indicated above. Ascent or descent that a vechicle should make is equal to the difference in the heights between the starting and the endpoint of a road segment, as shown in Fig.~\ref{fig:altitude}~\cite{gradient_wiki}.
Given the altitude and the length (hypotenuse) of the road segment, we can find the elevation. The measurement of all road gradient is with reference to the sea level. If the elevation data is available for every point on the globe, we can easily calculate the raise or drop in elevation between two points.

Digital Elevation Model (DEM)~\cite{zhang1994digital} is a digital representation of the Earth's topography. 3D Computer Graphics (3D-CG) represents the land surface generated from the terrain elevation data. DEM is created by collecting elevation from equally or unequally spaced points~\cite{remote_sensing}. The collected elevation information can be used for continuous spatial representation of any terrain. DEM just gives elevation information of terrain, including trees, or any human-made structures with reference to Mean Sea Level (MSL).

Depending upon how the elevation information is structured and stored, we can classify DEM into two types:
\begin{enumerate}
  \item Regular square grids,
  \item Triangulated irregular networks (TIN). 
\end{enumerate}
Grid DEM is stored as a rastor map in which each pixel has the information about the elevation of terrain. TIN data also includes features like peaks, slopes, conic pits. So, TIN-DEM considered being more accurate for extracting structural information. 
It is possible to obtain elevation information using field surveys from topographic contours, aerial photographs, or satellite imagery using photogrammetric techniques. Photogrammetry is an of three-dimensional measurements from two-dimensional data. Recently radar interferometric techniques and Laser altimetry have also been used to generate a DEM. Today very fine resolution DEMs at the near-global scale are readily available from various sources~\cite{remote_sensing}.

Some of the sources of global elevation data set are.
\begin{enumerate}
    \item Global 30 Arc-Second Elevation (GTOPO30): GTOPO~\cite{gtopo30} a global digital elevation model (DEM) with a horizontal grid spacing of 30 arc seconds (approximately 1 kilometer). GTOPO30, completed in late 1996, was developed over three years through a collaborative effort led by staff at the US Geological Survey’s EROS Data Center (EDC). It requires topographic information from several rasters and vector sources. 

    \item NOAA Global Land One-km Base Elevation (GLOBE): GLOBE~\cite{noaa-globe} is an internationally designed, developed, and independently reviewed global DEM, at a latitude-longitude grid spacing of 30 arc seconds (approx. 1km).
    
    \item Shuttle Radar Topography Mission (SRTM): SRTM~\cite{jarvis2008hole} is an international research effort that obtained DEM on a near-global scale to generate the most complete high-resolution digital topographic database of Earth. Using two radar antennas and a single pass, it collected sufficient data to create a digital elevation model. This 1-arc second global DEM has a spatial resolution of about 30 meters in most areas. Also, it covers most of the world with an absolute vertical height accuracy of 10 meters approximately.
    
    \item ASTER Global Digital Elevation Model: NASA and Japan’s joint operation was the birth of Advanced Spaceborne Thermal Emission and Reflection Radiometer (ASTER)~\cite{paul2004combining}. ASTER data are used to create detailed maps of the surface temperature of land, emissivity, reflectance, and elevation.
    
    \item Lidar DEM: It is an airplane laser system that sends light pulses. It reflects the terrain and the objects on it. LiDAR~\cite{gesch2009analysis} transmits and receives electromagnetic radiation of the near IR band. Unlike photogrammetry, LiDAR data collection is not affected by sun angle. LIDAR precision depends on GPS, IMU, and Laser measurements.
\end{enumerate}

\begin{table*}[htb]
    \begin{center}
      \begin{tabular}{ | m{2cm} | m{2cm} | m{2cm} | m{2cm} | }
        \hline
      \multicolumn{1}{|c||}{\textbf{DEM}} & \multicolumn{1}{c|}{\textbf{DEM Resolution}} & \multicolumn{1}{c|}{\textbf{DEM Accuracy}} & \multicolumn{1}{c|}{\textbf{Area of missing data}} \\
        \hline\hline
          \multicolumn{1}{|c||}{GTOPO30~\cite{global30arc}} & \multicolumn{1}{c|}{900 mts} & \multicolumn{1}{c|}{30m} & \multicolumn{1}{c|}{None} \\
        \hline
          \multicolumn{1}{|c||}{NOAA Globe~\cite{hastings1993global}} & \multicolumn{1}{c|}{900 mts} & \multicolumn{1}{c|}{30m} & \multicolumn{1}{c|}{None} \\
        \hline
          \multicolumn{1}{|c||}{SRTM~\cite{jarvis2008hole}} & \multicolumn{1}{c|}{90 mts and 30 mts} & \multicolumn{1}{c|}{10m} & \multicolumn{1}{c|}{Topographically steep areas} \\  
        \hline
          \multicolumn{1}{|c||}{ASTER GDEM~\cite{paul2004combining}} & \multicolumn{1}{c|}{30 mts} & \multicolumn{1}{c|}{14 mts} & \multicolumn{1}{c|}{Areas with constant cloud cover} \\
        \hline
          \multicolumn{1}{|c||}{Lidar~\cite{gesch2009analysis}} & \multicolumn{1}{c|}{1 mts} & \multicolumn{1}{c|}{15 cm} & \multicolumn{1}{c|}{available at certain areas} \\
        \hline
      \end{tabular}
    \end{center}
    \caption{Various DEM services}
        \label{tab:DEM_services}
\end{table*}

From the table~\ref{tab:DEM_services}, we can see that DEMs GTOPO30 and NOAA GLOBE has poor resolution and accuracy. LIDAR gives high accuracy and resolution, but it’s just for specialized enterprise coverage specific to certain areas. So, the choice of SRTM data is best for achieving good accuracy, and an acceptable resolution.

The data collected by SRTM has voids~\cite{robinson2014earthenv}. Two groups of scientists worked to fill the voids for different purposes, the CGIAR-CSI versions and the USGS HydroSHEDS dataset. The CGIAR-CSI version 4 provides the best global coverage for the full resolution SRTM dataset. The HydroSHEDS dataset was generated for hydrological applications and is suitable for consistent drainage and water flow information. In our investigation, we have used elevation information created by  Shuttle Radar Topography Mission (SRTM), which is provided by CGIAR-CSI. 

Using CGIAR-CSI GeoPortal, we can get the elevation data of terrain in the intervals of 5$^{\circ} \times 5^{\circ}$ tiles. The elevation produced by SRTM has a resolution of 90m at the equator. The elevation data are available in both ArcInfo ASCII and GeoTiff format~\cite{srtm_ascii_tiff}, which is easy to process for our purpose.

\begin{figure}[htb]
  \begin{center}
    \centering
    \includegraphics[scale=0.3]{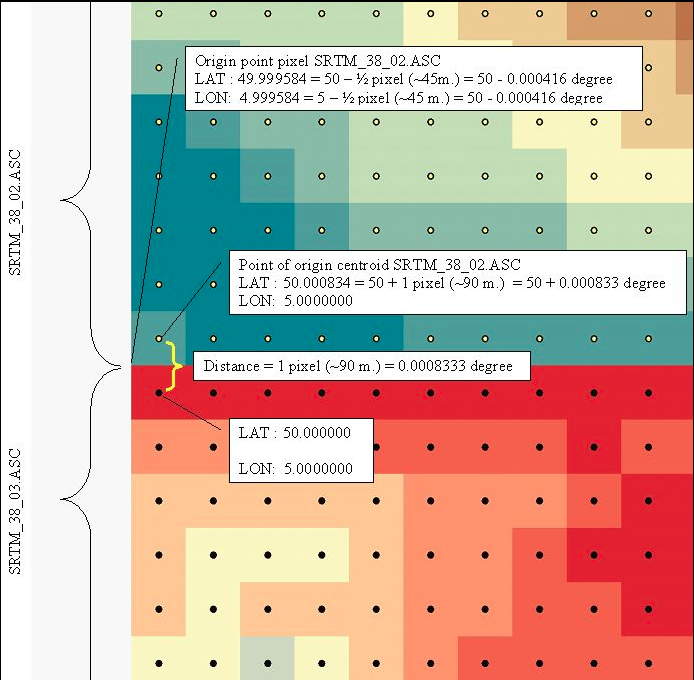}
    \caption{Information about dataset}
    \label{fig:srtm dataset}
  \end{center}
\end{figure}

Figure~\ref{fig:srtm dataset} gives an illustration of a sample SRTM data points.
The same vertical line points have the same longitude, but different latitudes will differ with 0.000833 degrees. Similarly, the same horizontal line points have the same latitude, but different longitudes differ by 0.000833 degrees. 

If we want the elevation at the Geo co-ordinate $P(x, y)$, we can find the nearest data point $ Q $ in the dataset and may assume $ Q $'s elevation as $P$'s elevation. It will possibly work well for city areas that are typically flat and have little elevations at a few places. However, it won't be accurate for mountainous roads. The elevation values differ a lot between two points on the stretch of road passing through mountains. So we need to generate a new data point using the dataset.

\subsubsection{Bilinear interpolation}

\begin{figure}[htb]
  \begin{center}
    \centering
    \includegraphics[scale=0.35]{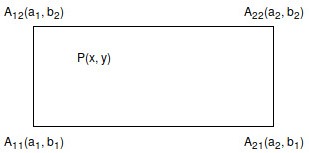}
    \caption{The four corners are the known data points and the point P at which we want to interpolate.}
    \label{fig:illustration_interpolation}
  \end{center}
\end{figure}

Generally, interpolation~\cite{davis1975interpolation} is used to find a new data point using the known data points. Fig~\ref{fig:illustration_interpolation} illustrates how interpolation is used.
Let the function $E:P\rightarrow R $, where $ R $ is the set of points on the real line, and $ P $ is at any point on the road. 
Suppose, we know the values of set of four given points $A_{11}$, $A_{21}$, $A_{22}$, $A_{12}$. 
First, we need to carry out linear interpolation along the $X$-direction that results:
$$ E(x, b_1) = \frac{a_2 - x}{a_2 - a_1} E(A_{11}) + \frac{x - a_1}{a_2 - a_1} E(A_{21}) $$
$$ E(x, b_2) = \frac{a_2 - x}{a_2 - a_1} E(A_{12}) + \frac{x - a_1}{a_2 - a_1} E(A_{22}) $$
Next, we need to perform linear interpolation along the $Y$-direction, that results:
$$ E(x, y) = \frac{b_2 - y}{b_2 - b_1} E(x, b_1) + \frac{y - b_1}{b_2 - b_1} E(x, b_2) $$
We can use non-linear interpolations also.

\subsubsection{Example}

Now let's consider following the mountainous road from Tirumala to Tirupati and compare the elevations along the path with naive, interpolation, and Google Maps data.
\begin{figure}[htb]
  \begin{center}
    \centering
    \includegraphics[scale=0.25,angle=-90]{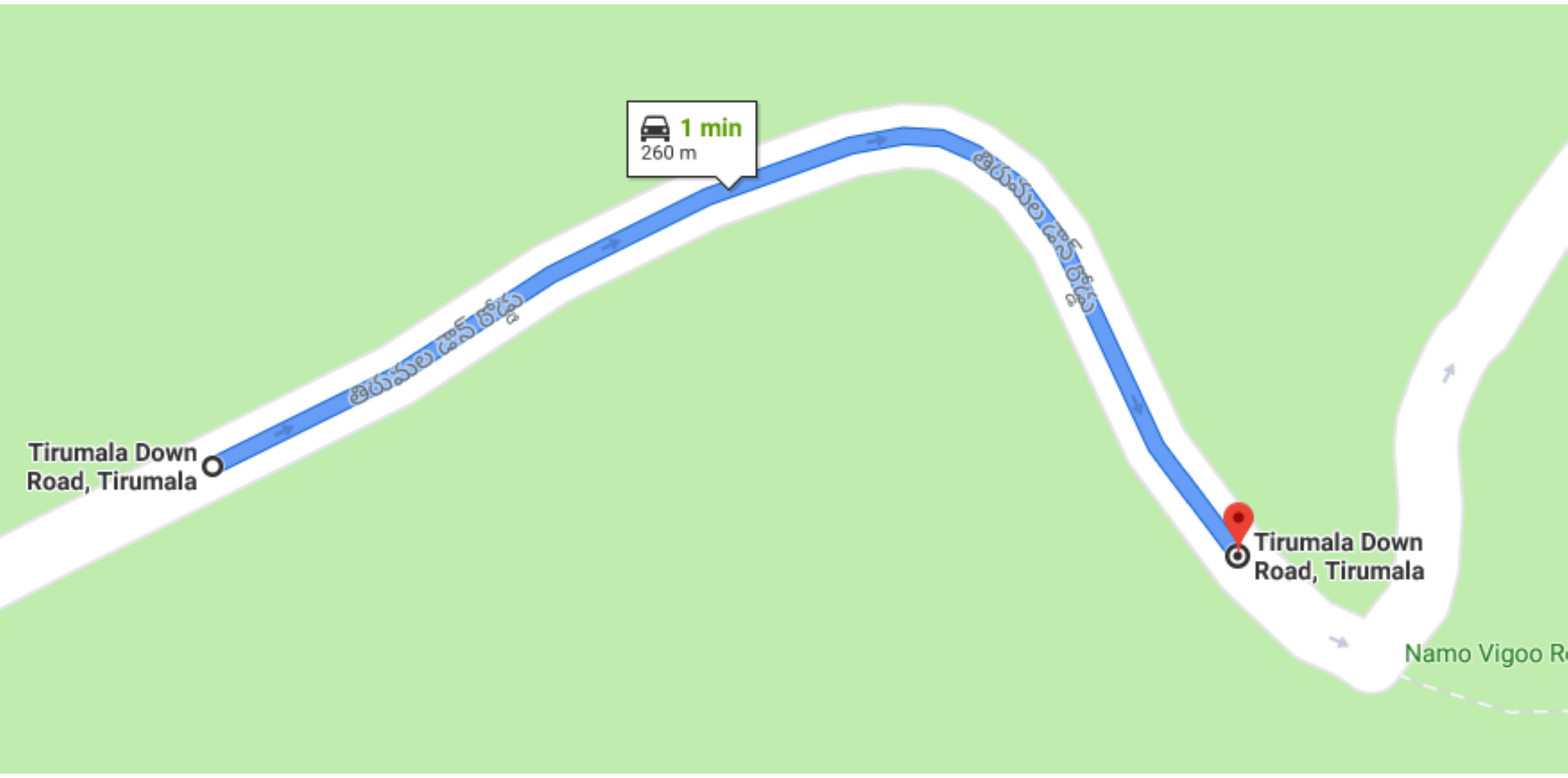}
    \caption{A stretch of 260m on a mountainous road.}
    \label{fig:ghat_road}
  \end{center}
\end{figure}
We first divide the path of 258m into seven segments. In the figure, the marker $ A $ is the starting position, and the marker $ H $ is the final position. 
$$A \xrightarrow{59m} B \xrightarrow{48m} C \xrightarrow{37m} D \xrightarrow{13m} E \xrightarrow{19m} F \xrightarrow{44m} G \xrightarrow{38m} H$$

\begin{table*}[htb]
    \begin{center}
      \begin{tabular}{ | m{4cm} | m{4cm} | m{5cm} | m{5cm} | }
        \hline
      \multicolumn{1}{|c||}{ \textbf{\backslashbox{Position}{Elevation}} } & \multicolumn{1}{c|}{\textbf{Rounding of to \newline nearest point(m)}} & \multicolumn{1}{c|}{\textbf{Interpolation}} & \multicolumn{1}{c|}{\textbf{Google Maps}} \\
        \hline\hline
          \multicolumn{1}{|c||}{A} & \multicolumn{1}{c|}{817} & \multicolumn{1}{c|}{822.33} & \multicolumn{1}{c|}{821.5} \\
        \hline
          \multicolumn{1}{|c||}{B} & \multicolumn{1}{c|}{839} & \multicolumn{1}{c|}{815.27} & \multicolumn{1}{c|}{812.22} \\
        \hline
          \multicolumn{1}{|c||}{C} & \multicolumn{1}{c|}{789} & \multicolumn{1}{c|}{809.46} & \multicolumn{1}{c|}{806.35} \\
        \hline
          \multicolumn{1}{|c||}{D} & \multicolumn{1}{c|}{800} & \multicolumn{1}{c|}{804.802} & \multicolumn{1}{c|}{804.167} \\
        \hline
          \multicolumn{1}{|c||}{E} & \multicolumn{1}{c|}{800} & \multicolumn{1}{c|}{806.41} & \multicolumn{1}{c|}{805.41} \\
        \hline
          \multicolumn{1}{|c||}{F} & \multicolumn{1}{c|}{800} & \multicolumn{1}{c|}{811.72} & \multicolumn{1}{c|}{810.55} \\
        \hline
          \multicolumn{1}{|c||}{G} & \multicolumn{1}{c|}{827} & \multicolumn{1}{c|}{822.99} & \multicolumn{1}{c|}{822.4} \\
        \hline
          \multicolumn{1}{|c||}{H} & \multicolumn{1}{c|}{827} & \multicolumn{1}{c|}{827.79} & \multicolumn{1}{c|}{827.1} \\
        \hline
      \end{tabular}
    \end{center}
    \caption{Comparing SRTM results with Google Maps}
        \label{tab:srtm_gmap}
\end{table*}

The table~\ref{tab:srtm_gmap} have the comparison of the SRTM elevation data without interpolation, with interpolation and Google Map elevation. From this, we can observe that SRTM data with interpolation gives us good accuracy.
Now we can calculate elevation at any given Geo co-ordinate. Therefore, we can now calculate the gradient of a road segment easily. Depending on the gradient, we calculate Road Gradient Factor.
Once the gradient of the road is known, we can use Equation~\ref{eq:FC} to change the edge weights of the graph into fuel consumption. Therefore, we can find the least fuel path for a given source and a destination. The next section deals with the results.
\section{Encoding the Polyline}

The route is fetched in the JSON format, where all the data are represented in the form of strings. Since the Geo coordinates are in float values, a precision of 5-6 decimal places with the integer part may require a string of hundreds of characters. 
For example, a single value of a latitude "13.351534" in JSON format requires nine bytes. However, any float can fit into 4 bytes. It implies that sending hundreds of points in JSON string format would lead to exchanging a lot of messages. The question is, can we apply some compression technique in encoding?

\subsection{A Simple Idea}

One can think of a simple-minded approach as follows. Suppose a float value is always considered as accurate up to five decimal places. Then we first convert a float value into an integer by multiplying it with $10^5$. Next, we convert the resulting integer into binary and then represent each byte as the corresponding character. However, some of these characters sometimes turn out to be control characters such as tab, space, etc. However, if we encode such a way that every character ASCII value between 63-126 are alphabets, brackets, etc., we can compress a polyline.

The Polyline Encoding provided in Algorithm~\ref{alg:PolylineEncode} help us to achieve it~\cite{PolylineEncoder}.
\begin{algorithm}
\SetKwFor{Proc}{Procedure}{}{end}
  \caption{Encoding Polyline}\label{alg:PolylineEncode}
  \Proc{Polyline\_Encoding($G$)}{ 
    // G-array of Geo coordinates (Polyline) \;
    String output \;
    factor $\leftarrow 10^5$\;
    end $\leftarrow$ length($G$)\;
    \For{each $i \in$ [0, end)}{
      // Encoding latitude of $i^{th}$ point\;
      output += \textbf{ENCODE}($G$[i][0]*factor)\; 
      // Encoding longitude of $i^{th}$ point\;
      output +=  \textbf{ENCODE}($G$[i][1]*factor)\; 
      }
      \Return{$output$}\;
  }
\end{algorithm}
\begin{algorithm}
\SetKwFor{Proc}{Procedure}{}{end}
\DontPrintSemicolon
  \caption{Encode a number}\label{alg:EncodeNumber}
  \Proc{Encode($A$)}{
     // A is an integer\:
     String output;\;
     $B$ = $A<<1$ ; // Left shift by 1 bit\;
    \If{A $<$ 0} {
      $B =\sim B$; // Complement\;
    }
    \While{B $\geq$ 0x20} {
     // Concatenation\;
      strcat(output, ASCII(0x20 $|$ ($B$ \& 0x1f)) + 63);\;
       $B>>5$; // Right shift by 5 bits\; 
    }
    \Return{output;}
  }
\end{algorithm}

The above polyline compression algorithm compresses JSON data of a route around 90kB to 15kB. When a user requests a route to our engine, it gives the response in JSON format. Since the response data is in compressed form, it helps to reduce network latency. To evaluate the effect of compression, we sent a hundred routes through the network to a user continuously. It took five seconds for an end-to-end transmission for non-compressed route data, and three seconds compressed where the number of hops between server and the user was one.

The above polyline compression algorithm compresses JSON data of a route around 90kB to 15kB. When a user requests a route to our engine, it gives the response in JSON format. Since the response data is in compressed form, it helps to reduce network latency. To evaluate the effect of compression, we sent a hundred routes through the network to a user continuously. It took five seconds for an end-to-end transmission for non-compressed route data, and three seconds compressed where the number of hops between server and the user was one.

    \section{Results and Analysis}
\label{ch:result}

In this section, we discuss the results from the Optimal Fuel Routing Machine (OFRM). At present,  OFRM web applicattion accepts Geo-coordinates or the addresses as source and destination.  The user can place markers for a source and another for a destination on the map. OFRM displays a complete description of the route between a given source-destination pair in terms of each road segment's distance, the associated directions, and the turns required. It also displays three values, namely, the distance, the time, and the fuel consumption needed to travel in that entire path. A few screenshots from OFRM are given below for the reader to understand the advantage of navigational outputs from OFRM.

\begin{figure*}[htb]
  \begin{center}
    \includegraphics[width=1.0\textwidth , height=6.5cm]{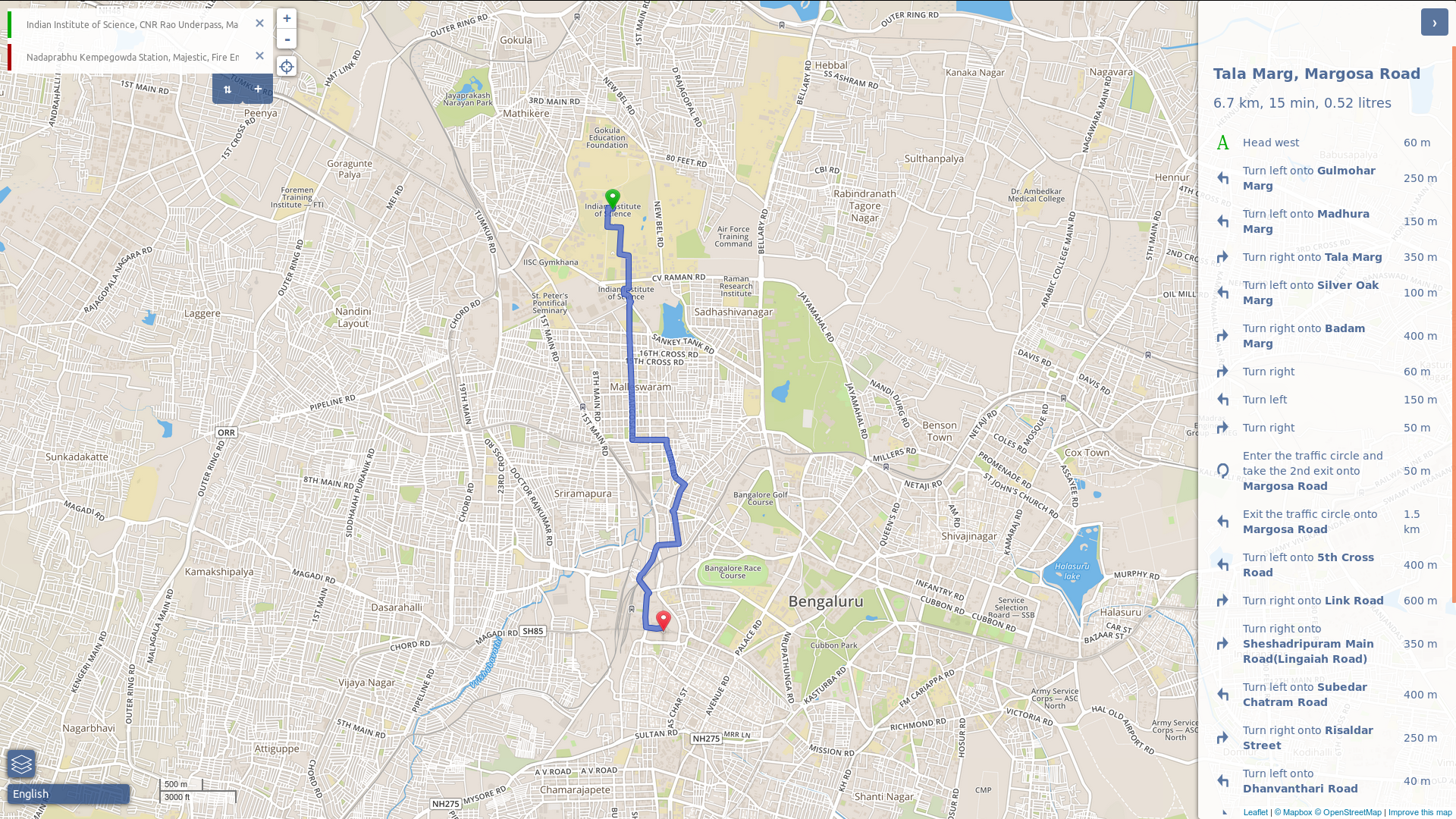}
  \end{center}
    \caption{IISC Bengaluru to Bengaluru Bus Station, input supplied as address}
    \label{fig:iisc_to_majestic_route}
\end{figure*}

From Fig~\ref{fig:iisc_to_majestic_route}, we can observe that there is a small box on the top left where the user specifies inputs for the source and the destination. A graphical display of the route between src to dst appears marked in thick blue. In the right part of the display window, the travel itinerary details are provided with exact navigational instructions for distance to be traveled for a segment of road and the type of turns (left, right or u-turns) to be executed. 
To show the map tiles, we used Mapbox APIs~\cite{mapbox}, for the detailed traveling itinerary, we used leaflet library~\cite{leaflet}.
The user may select the source and the destination on the map instead of providing the input address. The markers are associated with (lat, long) pairs. So, the marker-based inputs will be interpreted as Geo-coordinates for the routing engine. The user also has an option to supply the input directly as Geo-coordinates.

\begin{figure*}[htb]
  \begin{center}
    \includegraphics[width=1.0\textwidth , height=15cm]{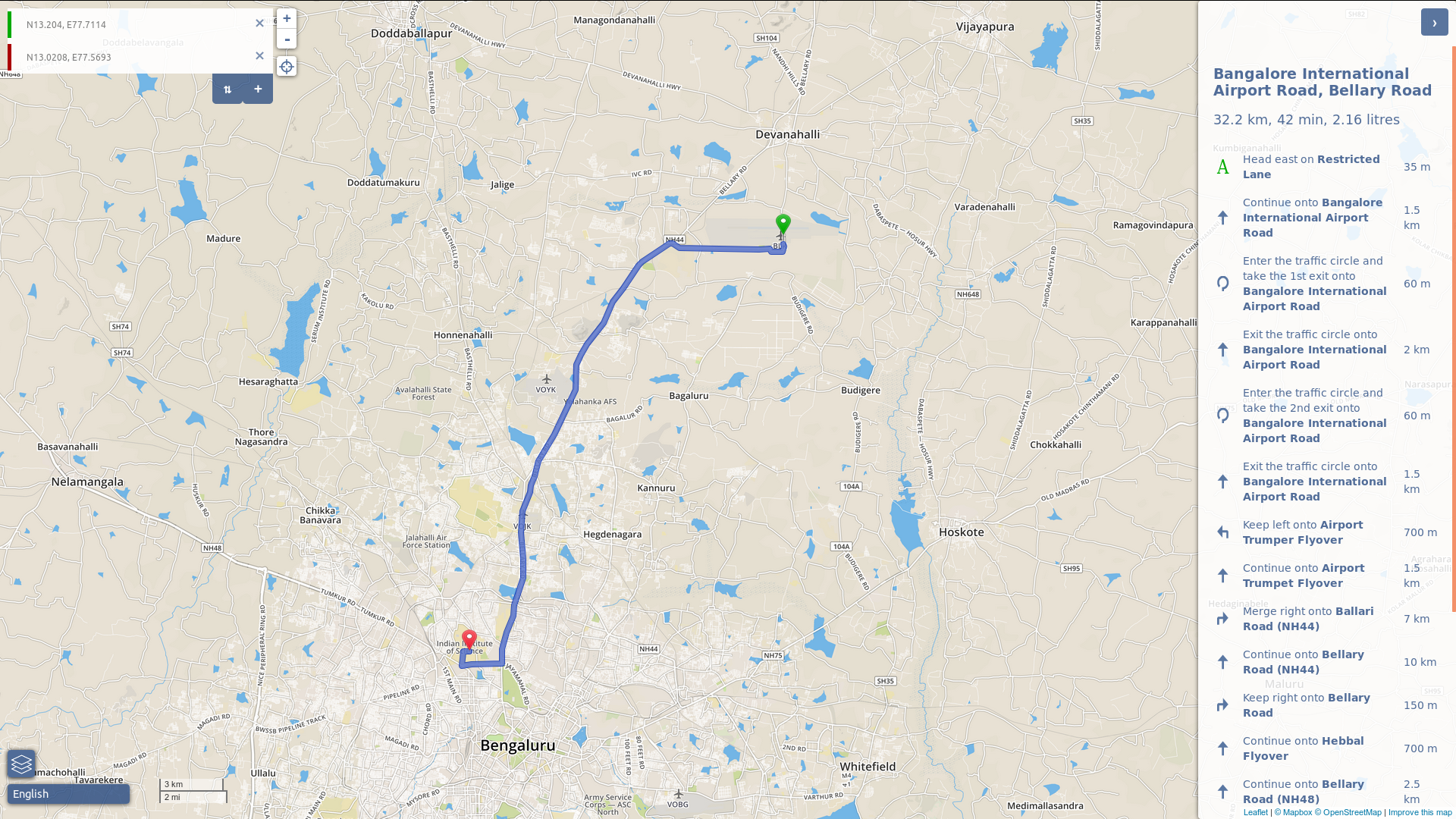}
  \end{center}
    \caption{Bengaluru Airport to IISC Bengaluru, selecting inputs on the map.}
    \label{fig:kia_to_iisc}
\end{figure*}

Fig~\ref{fig:kia_to_iisc} illustrates an example, where the user selected the source as Bengaluru's International Airport and the destination as IISC on the map. So, the input is treated as Geo-coordinates. As a result, we can see the graphical route and travel itinerary between the source and destination.

If a user wants to find a route that goes via some chosen places, then it is possible to request the routing engine to display a least fuel cost path by specifying any finite number of way points.
\begin{figure*}[htb]
  \begin{center}
    \includegraphics[width=1.0\textwidth , height=15cm]{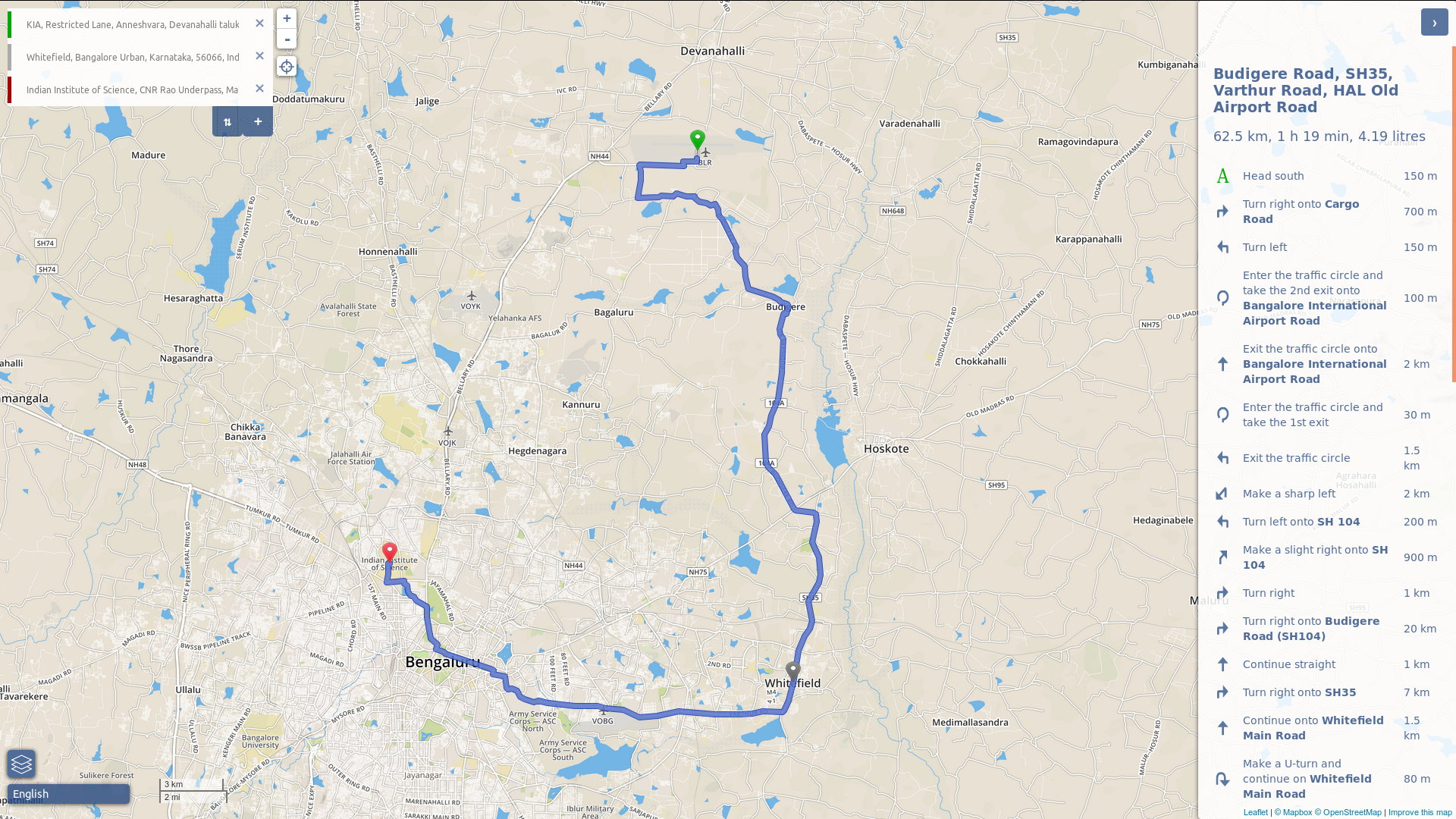}
  \end{center}
    \caption{Bengaluru Airport to IISC Bengaluru via Whitefield.}
    \label{fig:kia_whitefield_iisc}
\end{figure*}

For example, suppose a user wants to travel from the source (airport) to the destination(IISC) via Whitefield. Then the user gives the three inputs, the source, the destination, and the way point. The routing engine will give a route from the airport to IISC via Whitefield, which we can see in the Fig~\ref{fig:kia_whitefield_iisc}.

\subsubsection*{Comparing mountainous Roads}

To investigate how the road gradient affects fuel consumption, we selected some of the mountainous roads, namely,  Tirupati-Tirumala, Aluva-Munnar, Kullu-Manali, and Kalka-Shimla.

For each road, we compared the duration of the travel from Google Maps, the output from OFRM (our routing engine) without considering the Road Gradient, then the output of OFRM taking the gradient of the road from SRTM dataFactor. 

\begin{table*}[htb]
  \begin{center}
    \begin{tabular}{ | m{3cm} | m{3cm} | m{3cm} | m{3cm} | m{3cm} | }
     \hline
       one-way Rd & \multicolumn{1}{c|}{\textbf{Elevation}} & \multicolumn{1}{c|}{\textbf{Google Maps}} & \multicolumn{1}{c|}{\textbf{OFRM-Zero}} & \multicolumn{1}{c|}{\textbf{OFRM-Gradient}} \\
     \hline\hline
      \href{https://www.google.com/maps/dir/13.647243,79.4056888/13.6723405,79.351532/@13.6560224,79.3556799,14z/data=!3m1!4b1!4m6!4m5!2m3!6e0!7e2!8j1558402200!3e0}{Tirumala Up Rd - 17km} & \multicolumn{1}{c|}{710m gain} & \multicolumn{1}{c|}{32min}  & \multicolumn{1}{c|}{16min, 1.03 Litres} & \multicolumn{1}{c|}{29min, 1.58 Litres} \\
     \hline
      \href{https://www.google.com/maps/dir/13.6723405,79.351532/13.647243,79.4056888/@13.6598245,79.3684709,14z/data=!3m1!4b1!4m6!4m5!2m3!6e0!7e2!8j1558402200!3e0}{Tirumala Down Rd - 17km} & \multicolumn{1}{c|}{710m drop} & \multicolumn{1}{c|}{37min}  & \multicolumn{1}{c|}{21min, 1.13 Litres} & \multicolumn{1}{c|}{38min, 1.07 Litres} \\
    \hline
    \end{tabular}
    \end{center}
    
    \begin{center}
    \begin{tabular}{ | m{3cm} | m{3cm} | m{3cm} | m{3cm} | m{3cm} | }
     \hline
       two-way Rd & \multicolumn{1}{c|}{\textbf{Elevation}} & \multicolumn{1}{c|}{\textbf{Google Maps}} & \multicolumn{1}{c|}{\textbf{OFRM-Zero}} & \multicolumn{1}{c|}{\textbf{OFRM-Gradient}} \\
     \hline\hline
       \href{https://www.google.com/maps/dir/10.0450775,77.0486689/10.1525203,77.0764999/@10.0988009,77.0454722,13.25z/data=!4m6!4m5!2m3!6e0!7e2!8j1558323000!3e0}{Aluva - Munnar Rd 18km} & \multicolumn{1}{c|}{570m gain} & \multicolumn{1}{c|}{40min}  & \multicolumn{1}{c|}{29min, 1.31 Litres} & \multicolumn{1}{c|}{51min, 1.94 Litres} \\
     \hline
      Aluva - Munnar Rd 18km & \multicolumn{1}{c|}{570m drop} & \multicolumn{1}{c|}{40min}  & \multicolumn{1}{c|}{29min, 1.31 Litres} & \multicolumn{1}{c|}{51min, 1.94 Litres} \\
    \hline
    \end{tabular}
    \end{center}
    
    \begin{center}
    \begin{tabular}{ | m{3cm} | m{3cm} | m{3cm} | m{3cm} | m{3cm} | }
     \hline
       two-way Rd & \multicolumn{1}{c|}{\textbf{Elevation}} & \multicolumn{1}{c|}{\textbf{Google Maps}} & \multicolumn{1}{c|}{\textbf{OFRM-Zero}} & \multicolumn{1}{c|}{\textbf{OFRM-Gradient}} \\
     \hline\hline
        \href{https://www.google.com/maps/dir/31.9607863,77.1147839/32.2371711,77.1867908/@32.1033314,77.0623144,11.5z/data=!4m2!4m1!3e0}{Kullu to Manali Rd - 38km} & \multicolumn{1}{c|}{680m gain} & \multicolumn{1}{c|}{1hr7min}  & \multicolumn{1}{c|}{29min, 2.33 Litres} & \multicolumn{1}{c|}{43min, 2.85 Litres} \\
     \hline
       Manali to Kullu Rd - 38km & \multicolumn{1}{c|}{680m drop} & \multicolumn{1}{c|}{1hr05min}  & \multicolumn{1}{c|}{29min, 2.33 Litres} & \multicolumn{1}{c|}{43min, 2.85 Litres} \\
    \hline
    \end{tabular}
    \end{center}
    
    \begin{center}
    \begin{tabular}{ | m{3cm} | m{3cm} | m{3cm} | m{3cm} | m{3cm} | }
     \hline
       two-way Rd & \multicolumn{1}{c|}{\textbf{Elevation}} & \multicolumn{1}{c|}{\textbf{Google Maps}} & \multicolumn{1}{c|}{\textbf{OFRM-Zero}} & \multicolumn{1}{c|}{\textbf{OFRM-Gradient}} \\
     \hline\hline
       \href{https://www.google.com/maps/dir/31.0928073,77.1366113/30.9692217,77.1055441/@31.0311885,77.0792033,13z/data=!3m1!4b1!4m2!4m1!3e0}{Kalka - Shimla Rd - 25km} & \multicolumn{1}{c|}{520m gain} & \multicolumn{1}{c|}{40min}  & \multicolumn{1}{c|}{18min, 1.51 Litres} & \multicolumn{1}{c|}{36min, 1.84 Litres} \\
     \hline
      Kalka - Shimla Rd - 25km & \multicolumn{1}{c|}{520m drop} & \multicolumn{1}{c|}{40min}  & \multicolumn{1}{c|}{18min, 1.51 Litres} & \multicolumn{1}{c|}{36min, 1.84 Litres} \\
    \hline
    \end{tabular}
  \end{center}
  \caption{Comparison of several mountainous roads using Google Maps, our routing engine with and  without considering the gradient.}
  \label{tab:mountainous_roads}
\end{table*}
The above polyline compression algorithm compresses JSON data of a route around 90kB to 15kB. When a user requests a route to our engine, it gives the response in JSON format. Since the response data is in compressed form, it helps to reduce network latency. To evaluate the effect of compression, we sent a hundred routes through the network to a user continuously. It took five seconds for an end-to-end transmission for non-compressed route data, and three seconds compressed where the number of hops between server and the user was one.The different columns of Table~\ref{tab:mountainous_roads} represent the following values:
\begin{enumerate}
\item \textbf{Google Maps} - it gives the output from the routing engine of Google Maps. It gives the travel time without considering the traffic.
\item \textbf{OFRM-Zero} - it gives two outputs time and fuel consumed as estimated by the proposed routing engine without considering the gradient of the road.
\item \textbf{OFRM-Gradient} - it also gives two outputs time and the amount of fuel spent as estimated by the proposed routing engine when the road gradient is also considered. 
\end{enumerate}
From the Table~\ref{tab:mountainous_roads} we observe that output produced by \textit{OFRM-Gradient} appear more realistic in terms of the duration of travel time as well as the estimation of fuel consumption. Since the road gradient either ascent or descent slows down the vehicle speed and increases the total duration of travel time compared to the corresponding outputs obtained from {\em OFRM-zero}. 
\subsection{Mobile Application}

We also designed a mobile application using open source project osmbonuspack~\cite{osmbonus}, where the user can supply a source and a destination. The application sends a request to the OFRM routing engine and using the response from OFRM. The application gives a graphical display of the least fuel route between the source and the destination and the itinerary. 

The application can take the current location of the user from device's GPS chip as input. In this case, the user don't need to provide a source. The application displays a marker at the user's current location. As he changes the location, the marker location in the map also changes correspondingly. But we maintained the display of the route from his/her position to the destination. If the user goes outside the route that the routing engine provided, the mobile application sends a new request with the user's current position and gets corresponding route updates. 
\begin{figure}[htb]
    \begin{center}
    \begin{minipage}{0.5\linewidth}
    \begin{center}
        \includegraphics[scale=0.08]{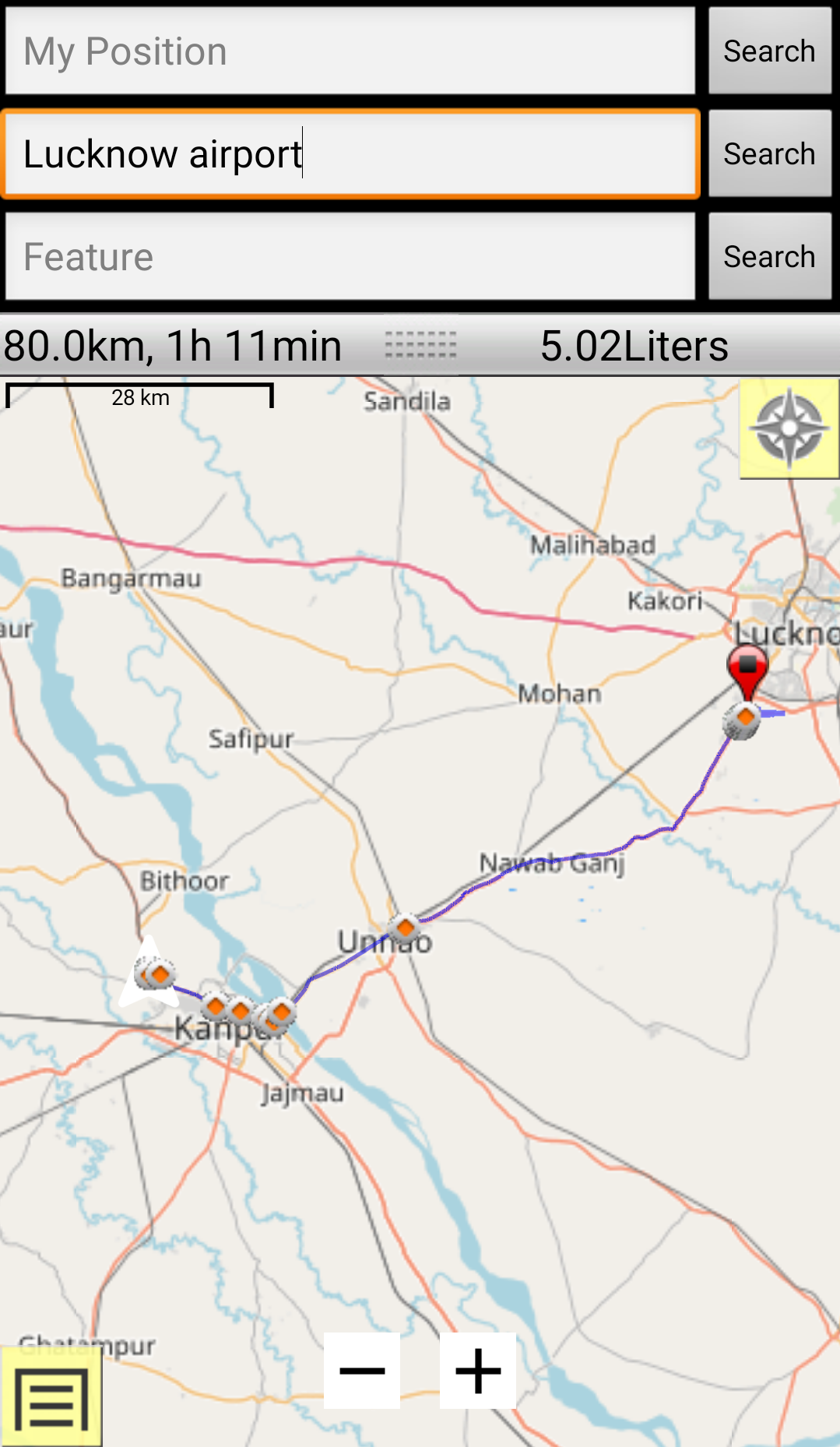}
        
        (a) Route at current location.
    \end{center}
        \end{minipage}\begin{minipage}{0.5\linewidth}
 \begin{center}
    \includegraphics[scale=0.08]{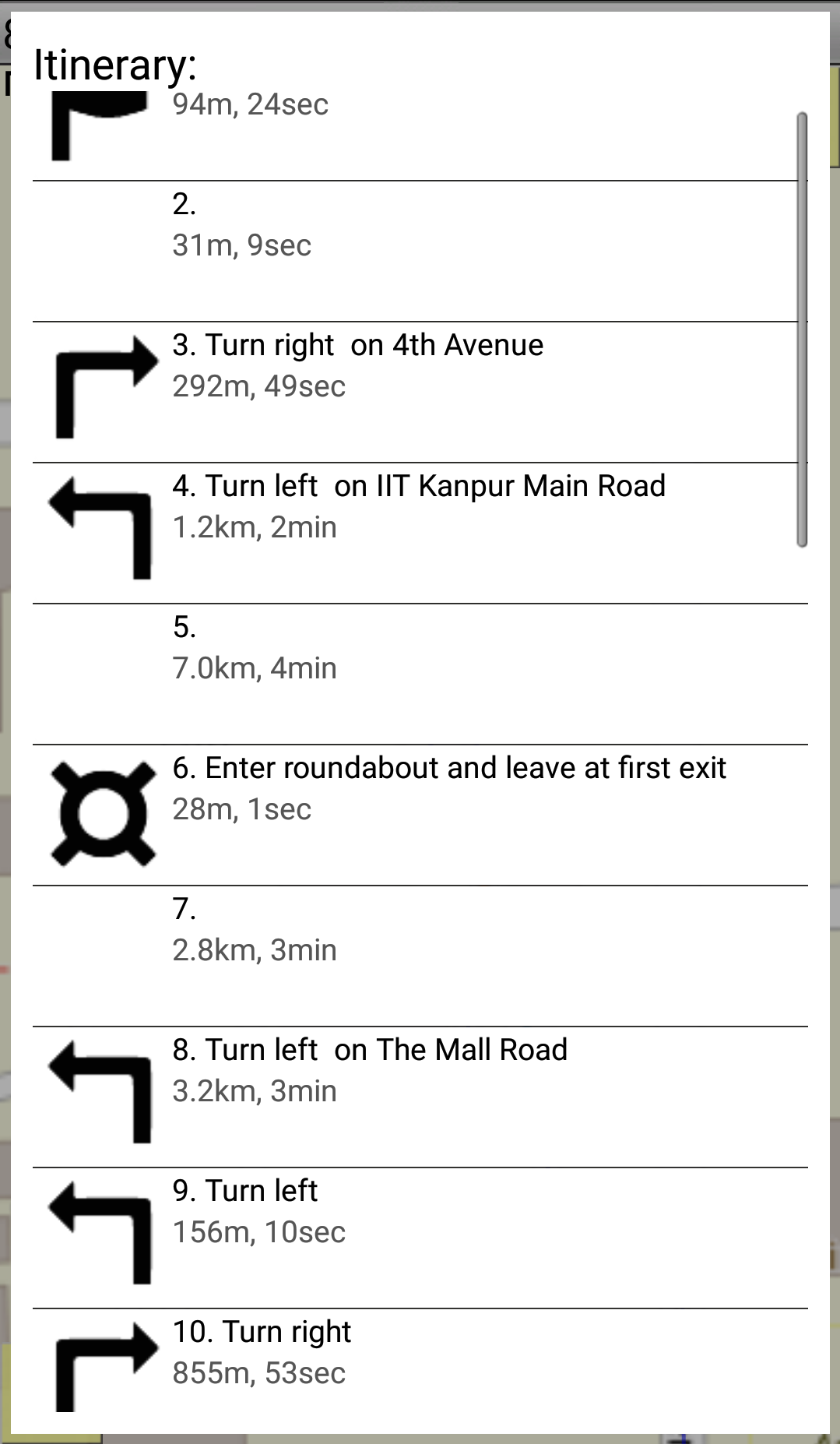} 
    
    (b) Travel itinerary.
 \end{center}
 \end{minipage}
    \end{center}
        \caption{Route and travel itinerary from current location of the user to destination.}
    \label{fig:ss_14}
\end{figure}
Figure~\ref{fig:ss_14} illustrate an example with GPS input.
On the left of the figure we can see the source as "My position", which is CSE Department and Destination as Lucknow Airport, we can see graphical route between source and destination, and also overall distance, duration and fuel required. In the right we have detailed travel itinerary.
\begin{figure}[htb]
    \begin{center}
        \begin{minipage}[b]{.5\linewidth}
        \centering\includegraphics[scale=0.08]{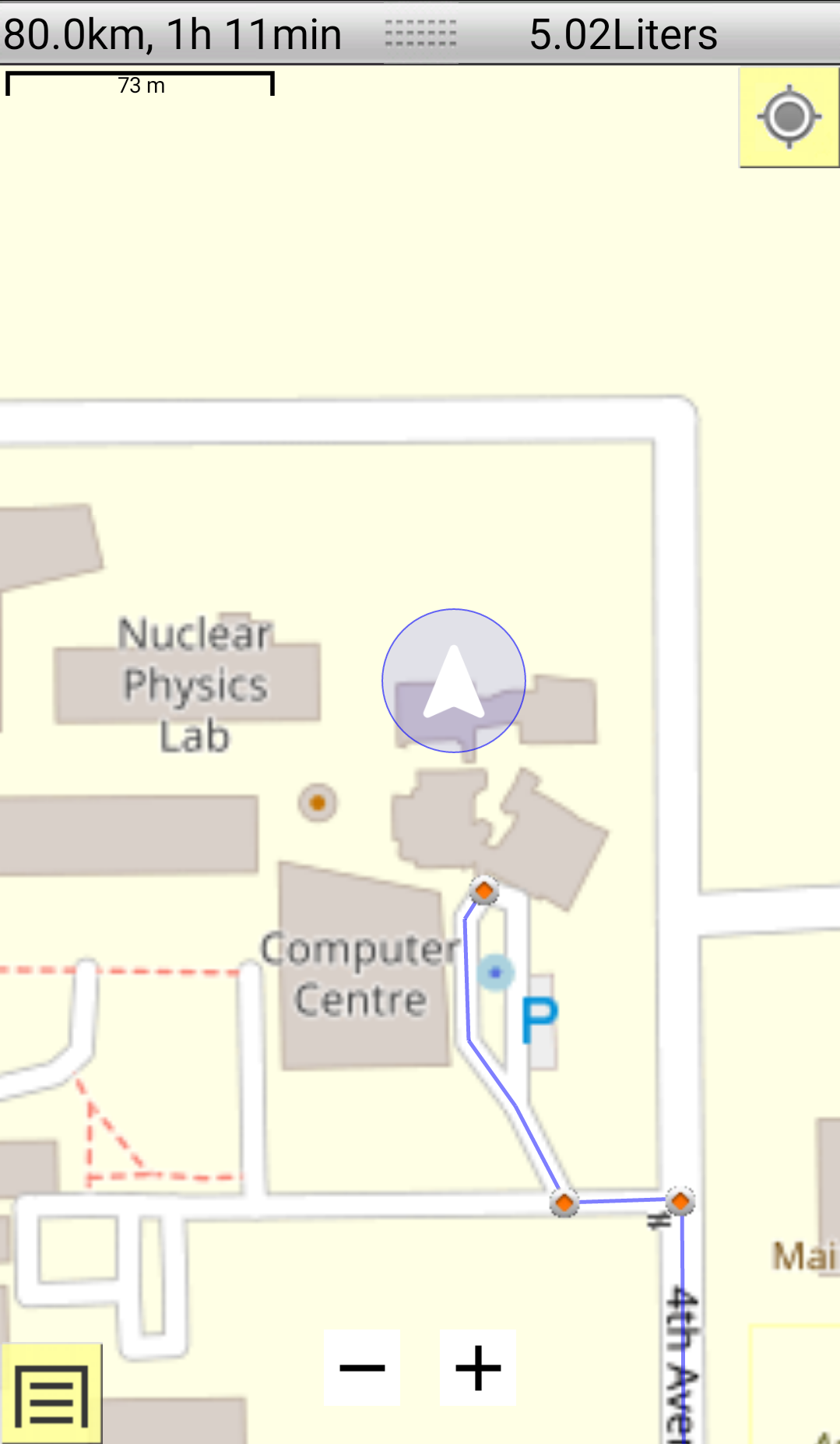}
        
        (a) Route at current location.
        \end{minipage}\begin{minipage}[b]{.5\linewidth}
         \centering\includegraphics[scale=0.08]{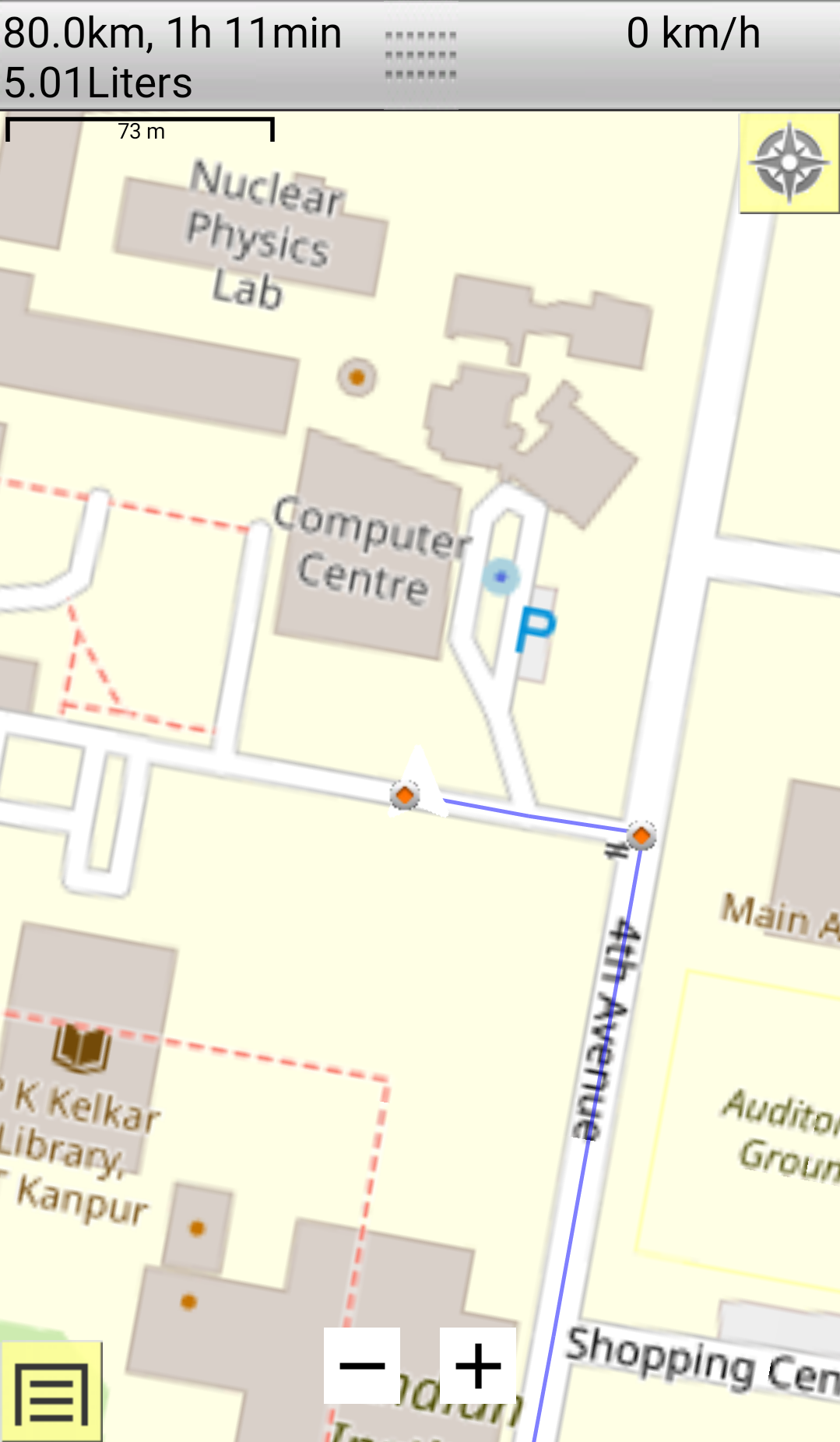}
         
         (b) Update with location. 
         \end{minipage}
    \end{center}
    \caption{As the current location changes the route gets updated.}
    \label{fig:ss_23}
\end{figure} 
Fig~\ref{fig:ss_23} illustrates the route update when the user moved away from the current route.  We can see the route from the current location of user (CSE Department) as the new route gets updated.

\subsection{Experiment on traffic trends in Indian cities}

Until now, we have chosen the vehicle velocity from Table~\ref{tab:OSRM_velocity} for the calculation of fuel consumption. The table specifies that the velocity is only dependent on the type of road. In other words, we used a set of time-invariant values for a vehicle's velocity. However, in a real-world scenario, human activities vary throughout the day, and traffic patterns vary accordingly. Choosing a constant velocity results in either underestimating or overestimating the fuel consumption. 

Choosing an appropriate value for the velocity is an important factor for computing fuel consumption. So we need to understand the trend of traffic flows throughout the day and obtain a dynamic set of velocity coefficients  for the velocities for the road types from Table~\ref{tab:OSRM_velocity}. In other words, by determining velocity coefficients it is possible to capture the dynamicity of the traffic conditions.  We approached the problem as follows in order to search for the appropriate velocity coefficients. We generated hundreds of source-destination pairs for different locations in five different cities across in India. Then using Google Map API, we determined the duration of travel  for each pair of source-destination on a weekday at different times of the day.

We compared the travel times between source-destination pairs at 9 AM, 5 PM, and 11 PM  with the travel time at 3 AM using.  At 9 AM and 5 PM, the vehicular movements reach peaks in almost all cities. At 11 PM, traffic is related to the nightlife in cities, but the traffic is expected to be relatively low compared to 9 AM or 5 PM. We have taken 3 AM outputs as the reference because minimal vehicular movements is expected around the time as it coincides with the normal cycle of human non-activity. Therefore, 3 AM is treated as zero traffic time. 
The values are tabulated in Table~\ref{tab:deviation_various_times}, where
t$_X$ represents the average time taken to travel 1km distance at "$X$:00" hours for $X \in \{3,9,17,23\}$.

\begin{table*}[htb]
    \begin{center}
      \begin{tabular}{ | m{5cm} | m{5cm} | m{5cm} | m{5cm} | }
        \hline
      \multicolumn{1}{|c||}{\textbf{city}} & \multicolumn{1}{c|}{\textbf{$\frac{t_9 - t_3}{t_3}$ }} & \multicolumn{1}{c|}{\textbf{$\frac{t_{17} - t_3}{t_3}$}} & \multicolumn{1}{c|}{\textbf{$\frac{t_{23} - t_3}{t_3}$}} \\
        \hline\hline
          \multicolumn{1}{|c||}{Bangaluru} & \multicolumn{1}{c|}{0.846 $\pm$ 0.232} & \multicolumn{1}{c|}{0.934 $\pm$ 0.213} & \multicolumn{1}{c|}{0.240 $\pm$ 0.035} \\
        \hline
            \multicolumn{1}{|c||}{Delhi} & \multicolumn{1}{c|}{0.698 $\pm$ 0.169} & \multicolumn{1}{c|}{0.670 $\pm$ 0.160} & \multicolumn{1}{c|}{0.242 $\pm$ 0.035} \\
        \hline
            \multicolumn{1}{|c||}{Hyderabad} & \multicolumn{1}{c|}{0.566 $\pm$ 0.153} & \multicolumn{1}{c|}{0.745 $\pm$ 0.148} & \multicolumn{1}{c|}{0.272 $\pm$ 0.055} \\
        \hline
            \multicolumn{1}{|c||}{Kolkata} & \multicolumn{1}{c|}{0.588 $\pm$ 0.140} & \multicolumn{1}{c|}{0.627 $\pm$ 0.193} & \multicolumn{1}{c|}{0.252 $\pm$ 0.042} \\
        \hline
            \multicolumn{1}{|c||}{Mumbai} & \multicolumn{1}{c|}{0.736 $\pm$ 0.278} & \multicolumn{1}{c|}{0.832 $\pm$ 0.170} & \multicolumn{1}{c|}{0.258 $\pm$ 0.044} \\
        \hline
      \end{tabular}
    \end{center}
    \caption{Additional time taken to travel 1km at different times of the day w.r.t 3AM.}
        \label{tab:deviation_various_times}
\end{table*}

From Table~\ref{tab:deviation_various_times}, we find that for travel in Bangaluru city:
\begin{itemize}
    \item At 9AM on an average it takes an extra time of (0.846 $\pm$ 0.232)t$_3$ mins to travel 1km distance,
    \item At 5PM on an average it takes an extra duration of (0.934 $\pm$ 0.213)t$_3$ mins, and
    \item At 11 PM on average, it takes an extra duration of (0.240 $\pm$ 0.035)t$_3$ mins.
\end{itemize}
Similarly, we can infer the factors for additional time requirements to account for traffic in the other cities, namely Mumbai, Kolkata, Delhi and Hyderabad. 


To test the accuracy of the proposed routing engine, we have compared its outputs for a set of sources-destination pairs with that of the Google Maps API when traveled around 9 AM, and we computed the ratio (average time provided by OFRM/average time provided by Google API). For Bangluru, the value of the ratio is  0.577, and it is 0.533 for Delhi. The reason for the huge deviation with respect to Google Maps is  attributed to the static velocity values used by OFRM without reference to variation in time of the day. Since Google Maps collects live data, its outputs are different from our routing engine outputs, which uses velocity values from the static Table~\ref{tab:OSRM_velocity}. It implies that one single static velocity table for all cities and all traffic trends cannot provide meaningful output.
So, we have to choose different set of velocity values for 9AM. Using statistics of average values from Table~\ref{tab:deviation_various_times}, we find that the average velocity of the vehicle at 3AM is almost twice compared to the velocity of the vehicle at 9AM. Using this relation we can generate velocity Table~\ref{tab:OFRM_velocity_traffic} for different roads for 9 AM, by taking Table~\ref{tab:OSRM_velocity} as reference.
\begin{table}[htb]
\begin{center}
  \begin{tabular}{ | m{5cm} | m{5cm} | }
    \hline
      \multicolumn{1}{|c|}{\textbf{Road Type}} & \multicolumn{1}{c|}{\textbf{velocity(km/hr)}} \\
    \hline\hline
      \multicolumn{1}{|c|}{motorway} & \multicolumn{1}{c|}{45} \\
    \hline
      \multicolumn{1}{|c|}{motorway\_link} & \multicolumn{1}{c|}{22.5} \\
    \hline
      \multicolumn{1}{|c|}{trunk} & \multicolumn{1}{c|}{42.5} \\
    \hline
      \multicolumn{1}{|c|}{trunk\_link} & \multicolumn{1}{c|}{20} \\
    \hline
      \multicolumn{1}{|c|}{primary} & \multicolumn{1}{c|}{32.5} \\
    \hline
      \multicolumn{1}{|c|}{primary\_link} & \multicolumn{1}{c|}{15} \\
    \hline
      \multicolumn{1}{|c|}{secondary} & \multicolumn{1}{c|}{27.5} \\
    \hline
      \multicolumn{1}{|c|}{secondary\_link} & \multicolumn{1}{c|}{12.5} \\
    \hline
      \multicolumn{1}{|c|}{tertiary} & \multicolumn{1}{c|}{20} \\
    \hline
      \multicolumn{1}{|c|}{tertiary\_link} & \multicolumn{1}{c|}{10} \\
    \hline
      \multicolumn{1}{|c|}{unclassified} & \multicolumn{1}{c|}{12.5} \\
    \hline
      \multicolumn{1}{|c|}{residential} & \multicolumn{1}{c|}{12.5} \\
    \hline
      \multicolumn{1}{|c|}{living\_street} & \multicolumn{1}{c|}{5} \\
    \hline
      \multicolumn{1}{|c|}{service} & \multicolumn{1}{c|}{7.5} \\
    \hline
  \end{tabular}
\end{center}
\caption{Velocity on various roads.}
    \label{tab:OFRM_velocity_traffic}
\end{table}

\begin{figure*}[htb]
    \begin{center}
\begin{subfigure}{.5\textwidth}
  \includegraphics[width=1.0\linewidth]{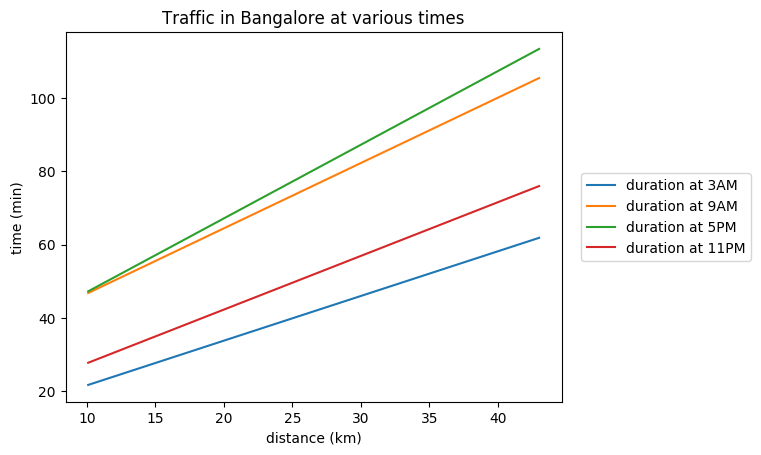}
  \label{fig:traffic_blore}
\end{subfigure}%
\begin{subfigure}{.5\textwidth}
  \includegraphics[width=1.0\linewidth]{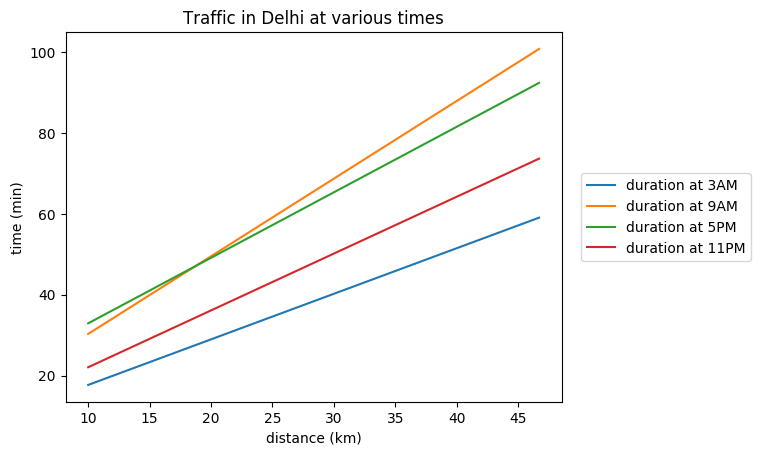}
  \label{fig:traffic_delhi}
\end{subfigure}

\begin{subfigure}{.5\textwidth}
  \includegraphics[width=1.0\linewidth]{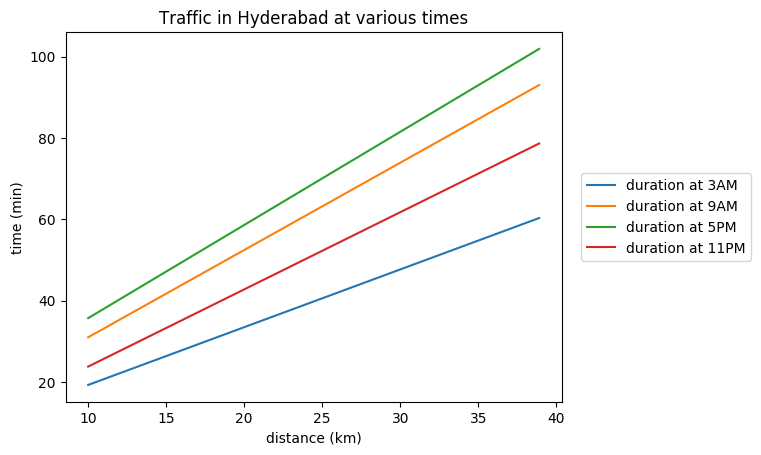}
  \label{fig:traffic_hyd}
\end{subfigure}%
\begin{subfigure}{.5\textwidth}
  \includegraphics[width=1.0\linewidth]{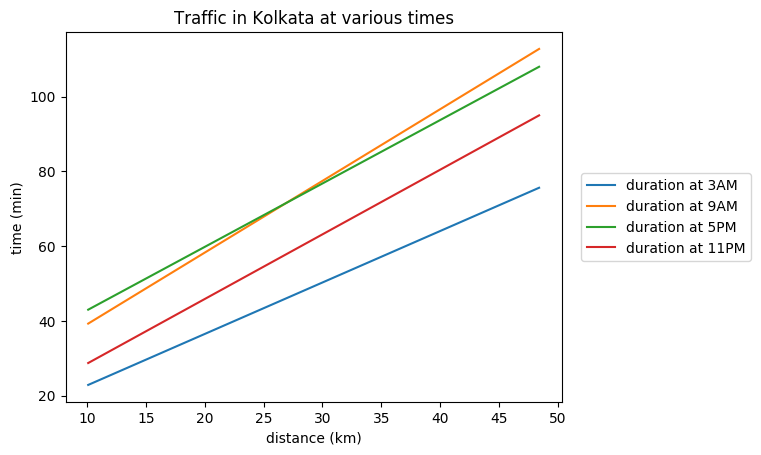}
  \label{fig:traffic_kolkata}
\end{subfigure}

\begin{subfigure}{.5\textwidth}
  \includegraphics[width=1.0\linewidth]{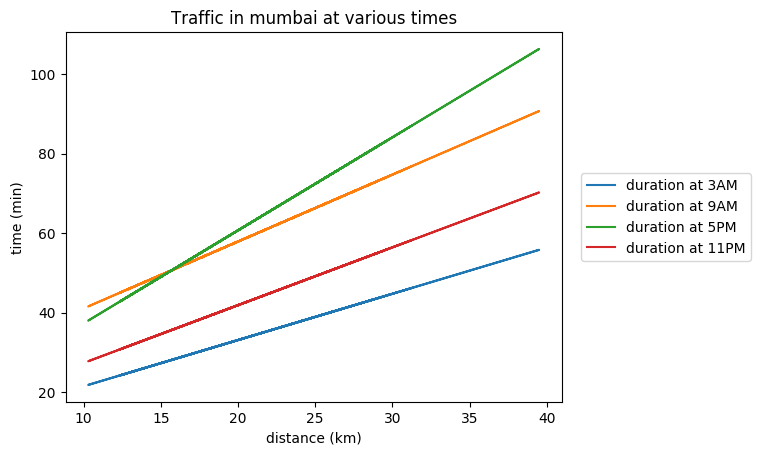}
  \label{fig:traffic_mumbai}
\end{subfigure}%
\end{center}
\caption{Analyzing traffic in several cities at various times.}
\label{fig:cities_traffic}
\end{figure*}

When we reevaluated our routing engine which uses velocity Table~\ref{tab:OFRM_velocity_traffic} with the Google Map at 9AM, the ratio (average time provided by OFRM/average time provided by Google API) in Bangluru increased to 0.886. In Delhi, it rose to 0.941. So it is essential to generate velocity values for more time intervals for different cities. OFRM can use these velocity values to produce results specific to the city’s traffic trends as in Fig.~\ref{fig:cities_traffic}. It clearly illustrates that the vehicular traffic during 9 PM and 5 PM have more or less similar trends. However, 11 AM traffic is a bit lower but not as low as 3 AM traffic.

 	\section{Conclusion}
\label{ch:conclusion}

We extended the  Open Source Routing Machine (OSRM) to build OFRM. OFRM provides the optimal fuel path between a given source-destination pair. It is possible to specify any number of waypoints between a source-destination pair. Using the proposed routing engine as the backend, one can design a pretty good user interface for road navigating. We have built a simple web application and mobile applications that use OSRM API in section~\ref{ch:result}. The OFRM routing engine's integration into a full-fledged navigation app like Google Maps or Open Street Maps is not difficult but will take time.

In particular, some of the future work are:
\begin{enumerate}
    \item We will focus on improving the methods of estimating fuel consumption to consider specific internal parameters of the car, namely vehicle radiator, exhaust heat, wind resistance, rolling resistance, and the car's acceleration, instead of using a statistical model like HERA.
    \item Offline mobile application - Instead of requesting a route to an online routing server, the routing engine should compute locally in the user’s mobile.
    \item We can include technology to convert itinerary travel data to speech, and also, we can improve the aesthetics of the user interface.
    \item We can add a feature, such that when a user selects an option of low in fuel, our routing engine should be able to provide a route via gas station.
\end{enumerate}

\section*{Acknowledgements}
This research is a part of the collaborative project on the eco-driving system with Hanyang University, Seoul, Republic of South Korea. It is sponsored by a grant from the Korean Evaluation Institute of Technology (KEIT).  We are thankful to KEIT for the grant. 


\balance
	\bibliographystyle{plain}
	\bibliography{sections/ref}

\end{document}